\documentclass[natbib]{svjour3}                     
\smartqed  
\usepackage{graphicx}
 \usepackage{aps-bibstyle}  
\newcommand{\keV}{\rm{\, keV }}
\newcommand{\MeV}{\rm{\, MeV }}

\newcommand{\erg}{\rm{\, erg }}

\newcommand{\beq}{\begin{equation}}
\newcommand{\eeq}{\end{equation}}
\newcommand{\ba}{\begin{array}}
\newcommand{\ea}{\end{array}}

\newcommand{\D}{{\mathcal {D}}}

\newcommand{\cjaa}{{\it Chinese~J.~Astron.~Astrophys.}}

\newcommand{\prd}   {{\it Physical Review D}}
\newcommand{\pre}   {{\it Physical Review E}}

\newcommand{\aj}{\mbox{\it Astronomical J.}}
\newcommand{\apj}{\mbox{\it Astrophys. J.}}
\newcommand{\apjl}{\mbox{\it Astrophys. J.}}
\newcommand{\apjs}{\mbox{\it Astrophys. J.}}
\newcommand{\apss}{\mbox{\it Astrophys. ans Space Science}}
\newcommand{\aap}{\mbox{\it Astron. Astrophys.}}
\newcommand{\aaps}{\mbox{\it Astron. Astrophys. Supp.}}
\newcommand{\araa}{\mbox{\it Annu. Rev. Astron. Astrophys.}}

\newcommand{\jcap}{\mbox{\it J. Cosmol. Astropart. Phys.}}
\newcommand{\mnras}{\mbox{\it Mon. Not. R. Astron. Soc.}}
\newcommand{\nar}{\mbox{\it New Astronomy Rev.}}
\newcommand{\nat}{\mbox{\it Nature}}
\newcommand{\na}{\mbox{\it New A.}}

\newcommand{\physrep}{\mbox{\it Phys. Rep.}}

\newcommand{\planss}{\mbox{\it Planetary and Space Science}}
\newcommand{\ssr}{\mbox{\it Space Science Rev.}}
\def\lsim{\raisebox{-0.3ex}{\mbox{$\stackrel{<}{_\sim} \,$}}}
\def\gsim{\raisebox{-0.3ex}{\mbox{$\stackrel{>}{_\sim} \,$}}}

\begin{document}

\title{Physics of Gamma-Ray Bursts Prompt Emission
}


\author{Asaf Pe'er 
}


\institute{A. Pe'er \at
              Physics Department, University College Cork, Cork, Ireland \\
              Tel.: +353-21-4902594\\
              \email{a.peer@ucc.ie}          
}

\date{Received: date / Accepted: date}

\maketitle

\begin{abstract}
In recent years, our understanding of gamma-ray bursts (GRB) prompt
emission has been revolutionized, due to a combination of new
instruments, new analysis methods and novel ideas. In this review, I
describe the most recent observational results and the current
theoretical interpretation. Observationally, a major development is
the rise of time-resolved spectral analysis. These led to (I)
identification of a distinguished high energy component, with GeV
photons often seen at a delay; and (II) firm evidence for the
existence of a photospheric (thermal) component in a large number of
bursts.  These results triggered many theoretical efforts aimed at
understanding the physical conditions in the inner jet regions from
which the prompt photons are emitted, as well as the spectral
diversity observed.  I highlight some areas of active theoretical
research. These include: (I) understanding the role played by magnetic
fields in shaping the dynamics of GRB outflow and spectra; (II)
understanding the microphysics of kinetic and magnetic energy
transfer, namely accelerating particle to high energies in both shock
waves and magnetic reconnection layers; (III) understanding how
sub-photospheric energy dissipation broadens the ``Planck'' spectrum;
and (IV) geometrical light aberration effects. I highlight some of
these efforts, and point towards gaps that still exist in our
knowledge as well as promising directions for the future.

\end{abstract}

\section{Introduction}
\label{intro}

In spite of an extensive study for nearly a generation, understanding
of gamma-ray bursts (GRB) prompt emission still remains an open
question. The main reason for this is the nature of the prompt
emission phase: the prompt emission lasts typically a few seconds (or
less), without repetition and with variable lightcurve. Furthermore,
the spectra vary from burst to burst, and do not show any clear
feature that could easily be associated with any simple emission
model. This is in contrast to the afterglow phase, which lasts much
longer, up to years, with (relatively) smooth, well characteristic
behavior. These features enable afterglow studies using long term,
multi-waveband observations, as well as relatively easy comparison
with theories.

Nonetheless, I think it is fair to claim that in recent years
understanding of GRB prompt emission has been revolutionized. This
follows the launch of {\it Swift} satellite in 2004 and {\it Fermi}
satellite in 2008. These satellites enable much more detailed studies
of the prompt emission, both in the spectral and temporal domains. The
new data led to the realization that the observed spectra are composed
of several distinctive components. (I) A thermal component identified
on top of a non-thermal spectra was observed in a large number of
bursts. This component show a unique temporal behavior. (II) There are
evidence that the very high energy ($>$~GeV) part of the spectra
evolve differently than the lower energy part, hence is likely to have
a separate origin. (III) The sharp cutoff in the lightcurves of many
GRBs observed by {\it Swift} enables a clear discrimination between
the prompt and the afterglow phases.

The decomposition of the spectra into separate components, presumably
with different physical origin, enabled an independent study of the
properties of each component, as well as study of the complex
connection between the different components. Thanks to these studies,
we are finally reaching a critical point in which a self consistent
physical picture of the GRB prompt emission, more complete than ever
is emerging.  This physical insight is of course a crucial link that
connects the physics of GRB progenitor stars with that of their
environments.

Many of the ideas gained in these studies are relevant to many other
astronomical objects, such as active galactic nuclei (AGNs), X-ray
binaries (XRBs) and tidal disruption events (TDEs). All these
transient objects share the common feature of having (trans)-
relativistic jetted outflows. Therefore, despite the obvious
differences, many similarities between various underlying physical
processes in these objects and in GRBs are likely to exist. These
include the basic questions of jet launching and propagation, as well
as the microphysics of energy transfer via magnetic reconnection and
particle acceleration to high energies. Furthermore, understanding the
physical conditions that exist during the prompt emission phase
enables the study of other fundamental questions such as whether GRBs
are sources of (ultra-high energy) cosmic rays and neutrinos, as well
as the potential of detecting gravitational waves associated with
GRBs.

In this review, I will describe the current (Dec. 2014) observational
status, as well as the emerging theoretical picture. I will emphasis a
major development of recent years, namely the realization that
photospheric emission may play a key role, both directly and
indirectly, as part of the observed spectra.  I should stress though
that in spite of several major observational and theoretical
breakthroughs that took place in recent years, our understanding is
still far from being complete. I will discuss the gaps that still
exist in our knowledge, and novel ideas raised in addressing them. I
will point to current scientific efforts, which are focused on
different, sometimes even perpendicular directions.

The rapid progress in this field is the cause of the fact that in the
past decade there have been very many excellent reviews covering
various aspects of GRB phenomenology and physics. A partial list
includes reviews by \cite{Waxman03, Piran04, ZM04, Meszaros06, Nakar07, Zhang07,
  FP08, GRF09, AB11, GM12, Bucciantini12, GR13, Daigne13, Zhang14,
  KZ14, Berger14, MR14}. My goal here is not to compete with these
reviews, but to highlight some of the recent - partially, still
controversial results and developments in this field, as well as to
point into current and future directions which are promising paths.

This review is organized as follows. In section \ref{sec:observations}
I discuss the current observational status. I discuss the lightcurves
(\S\ref{sec:lightcurve}), observed spectra (\S\ref{sec:spectra}),
polarization (\S\ref{sec:polarization}), counterparts at high and low
energies (\S\ref{sec:wavebands}) and notable correlations
(\S\ref{sec:correlations}). I particularly emphasis the different
models used today in fitting the prompt emission spectra. Section
\ref{sec:theory} is devoted to theoretical ideas. To my opinion, the
easiest way to understand the nature of GRBs is to follow the
various episodes of energy transfer that occur during the GRB
evolution. I thus begin by discussing models of GRB progenitors (\S
\ref{sec:progenitors}), that provide the source of energy. This
follows by discussing models of relativistic expansion, both
``hot'' (photon-dominated)  (\S \ref{sec:expansion}), and
``cold'' (magnetic-dominated) (\S \ref{sec:magnetized_flow}). I then
discuss recent progress in understanding how dissipation of the
kinetic and/or magnetic energy is used in accelerating
particles to high-energies (\S \ref{sec:acc}). I complete with the
discussion of the final stage of energy conversion - namely, radiative
processes by the hot particles as well as the photospheric
contribution (\S \ref{sec:radiation}), which lead to the observed signal. 
I conclude with a look into the future in \S\ref{sec:summary}.

\section{Key observational properties}
\label{sec:observations}

\subsection{Lightcurves}
\label{sec:lightcurve}
\begin{figure}
\includegraphics[width=13cm]{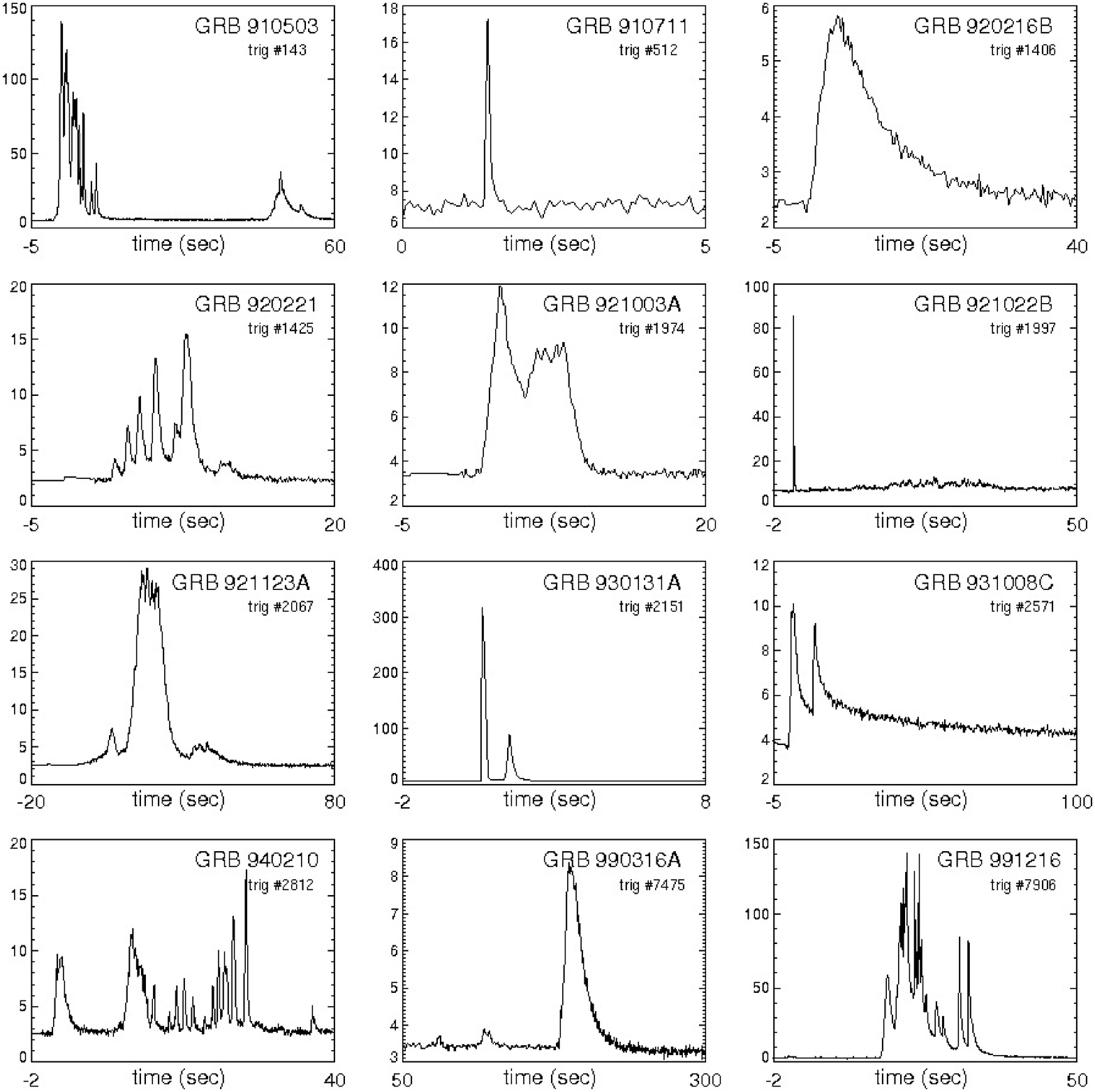}
\caption{
Light curves of 12 bright gamma-ray bursts detected by
BATSE. Gamma-ray bursts light curves display a tremendous amount of
diversity and few discernible patterns. This sample includes short
events and long events (duration ranging from milliseconds to
minutes), events with smooth behavior and single peaks, and events
with highly variable, erratic behavior with many peaks. Created by
Daniel Perley with data from the public BATSE archive
(http://gammaray.msfc.nasa.gov/batse/grb/catalog/).
}
\label{fig:lightcurves}
\end{figure}

The most notable property of GRB prompt emission lightcurve is that it
is irregular, diverse and complex. No two gamma-ray bursts lightcurves
are identical, a fact which obviously makes their study
challenging. While some GRBs are extremely variable with variability
time scale in the millisecond range, others are much smoother. Some
have only a single peak, while others show multiple peaks; see Figure
\ref{fig:lightcurves}. Typically, individual peaks are not symmetric,
but show a ``fast rise exponential decay'' (FRED) behavior.

The total duration of GRB prompt emission is traditionally defined by
the ``$T_{90}$'' parameter, which is the time interval between the
epoch when 5\% and 95\% of the total fluence is detected. As
thoroughly discussed by [\cite{KZ14}], this (arbitrary) definition is very
subjective, due to many reasons. (1) It depends on the
energy range and sensitivity of the different detectors; (2) Different
intrinsic lightcurves - some lightcurves are very spiky with gaps
between the spikes, while others are smooth; (3) No discrimination is
made between the ``prompt'' phase and the early afterglow emission; (4)
It does not take into account the difference in redshifts between the
bursts, which can be substantial.

\begin{figure}
\includegraphics[width=13cm]{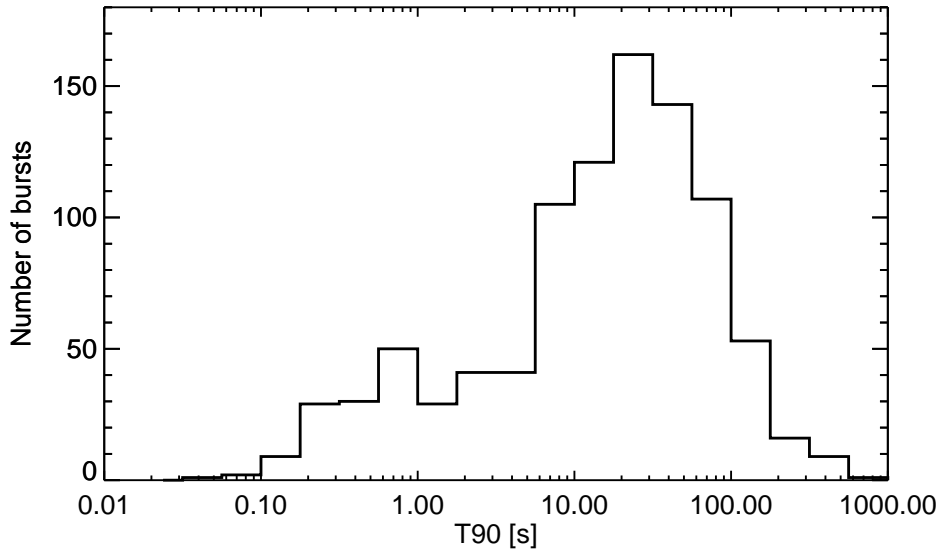}
\caption{Distribution of GRB durations ($T_{90}$) of 953 bursts in the
  Fermi-GBM (50 - 300 keV energy range).  Taken from the 2nd Fermi
  catalog, [\cite{VonKienlin+14}]. 159 (17\%) of the bursts are
  ``short''.  }
\label{fig:T90}
\end{figure}

In spite of these drawbacks, $T_{90}$ is still the most commonly used
parameter in describing the total duration of the prompt phase.
While $T_{90}$ is observed to vary between milliseconds and thousands
of seconds (the longest to date is GRB111209A, with duration of $\sim
2.5 \times 10^4$~s [\cite{Gendre+13}]), from the early 1990's, it was
noted that the $T_{90}$ distribution of GRB's is bimodal
[\cite{Kouveliotou+93}]. About $\lsim 1/4$ of GRBs in the BATSE
catalog are ``short'', with average $T_{90}$ of $\sim 0.2 - 0.3$~s, and
roughly $3/4$ are ``long,'' with average $T_{90} \approx 20-30$~s
[\cite{Paciesas+99}]. The boundary between these two distributions is at
$\sim 2$~s. Similar results are obtained by {\it Fermi} (see Figure
\ref{fig:T90}), though the subjective definition of $T_{90}$ results
in a bit different ratio, where only 17\% of Fermi-GBM bursts are
considered as ``short'', the rest being long [\cite{Paciesas+12, Qin+13,
  VonKienlin+14}]. Similar conclusion - though with much smaller
sample, and even lesser fraction of short GRBs are observed in the
{\it Swift}- Bat catalog [\cite{Sakamoto+11}] and by {\it Integral}
[\cite{Bosnjak+14}]. These results do not change if instead one uses
$T_{50}$ parameter, defined in a similar way.

These results are accompanied by different hardness ratio (the ratio
between the observed photon flux at the high and low energy bands of
the detector), where short bursts are, on the average harder (higher
ratio of energetic photons) than long ones
[\cite{Kouveliotou+93}]. Other clues for different origin are the
association of only the long GRBs with core collapse supernova, of type Ib/c
[\cite{Galama+98, Hjorth+03, Stanek+03, Campana+06, Pian+06, Cobb+10,
    Starling+11}] which are not found in short GRBs [\cite{Kann+11}];
association of short GRBs to galaxies with little star-formation (as
opposed to long GRBs which are found in star forming galaxies) and
residing at different locations within their host galaxies than long
GRBs [\cite{Gehrels+05, Fox+05, Villasenor+05, Barthelmy+05,
    Hjorth+05, Bloom+06, Troja+08, FB13}].  Altogether, these results
thus suggest two different progenitor classes. However, a more careful
analysis reveals a more complex picture with many outliers to these
rules [e.g., \cite{Zhang+07, NFP09, Zhang+09, Virgili+11, Norris+11,
    Berger11, Bromberg+13, Fong+13}]. It is therefore possible - maybe
even likely - that the population of short GRBs may have more than a
single progenitor (or physical origin).  In addition, there have been
several claims for a small, third class of ``intermediate'' GRBs, with
$T_{90} \sim 2$~s [\cite{Mukherjee+98, Horvath98, Horvath+06,
    Veres+10}], but this is still controversial [e.g.,
  \cite{Hakkila+03, Bromberg+13}].

\begin{figure}
\includegraphics[width=12cm]{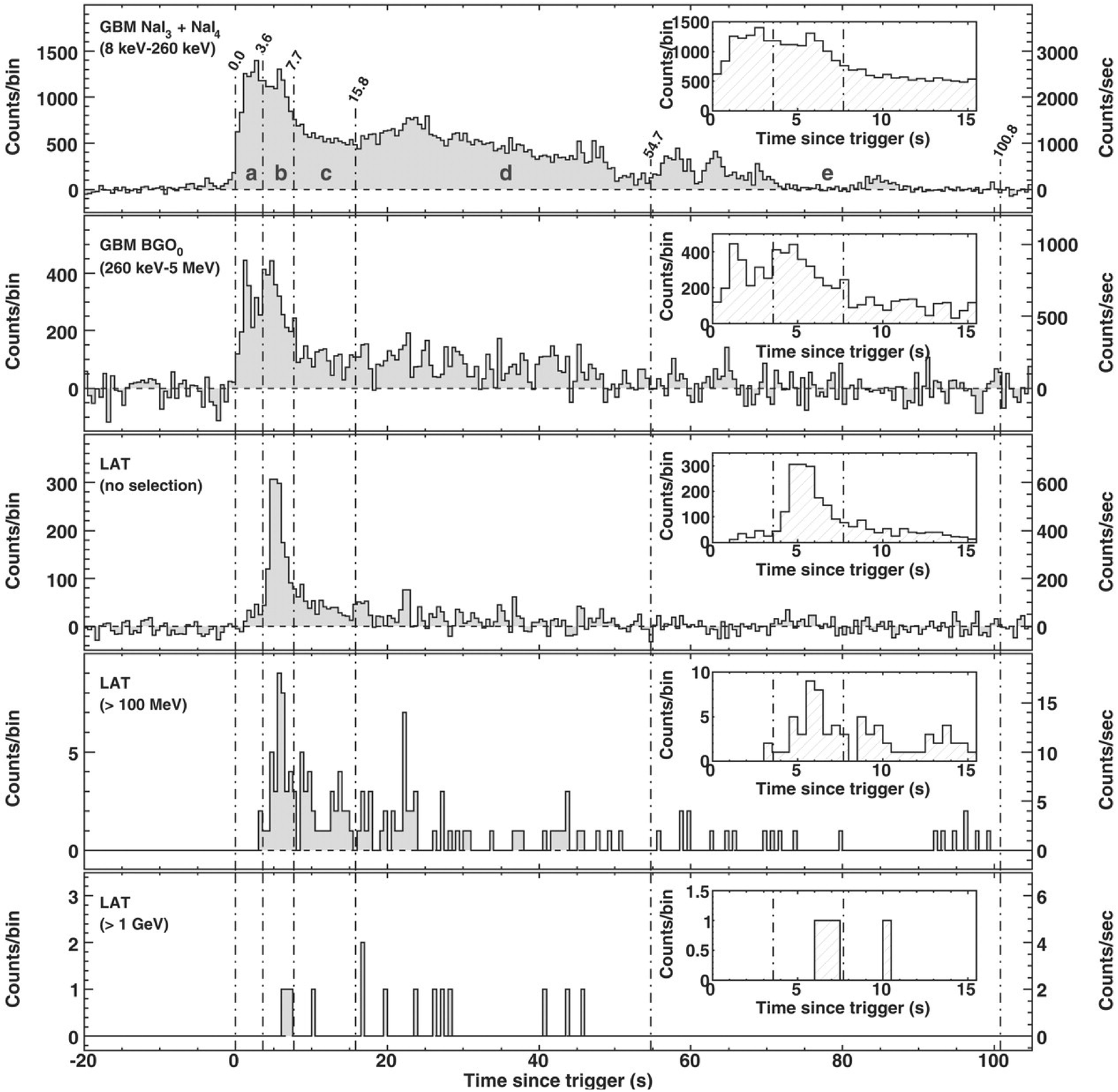}
\caption{ Light curves for GRB 080916C observed with the GBM and the
  LAT detectors on board the {\it Fermi} satellite, from lowest to
  highest energies. The top graph shows the sum of the counts, in the 8- to 260-keV
  energy band, of two NaI detectors. The second is the corresponding
  plot for BGO detector 0, between 260 keV and 5 MeV. The third shows
  all LAT events passing the on board event filter for
  gamma-rays. (Insets) Views of the first 15 s from the trigger
  time. In all cases, the bin width is 0.5 s; the per-second counting
  rate is reported on the right for convenience. Taken from
  \cite{Abdo+09b}.  }
\label{fig:080916c}
\end{figure}

To further add to the confusion, the lightcurve itself vary with
energy band (e.g., Figure \ref{fig:080916c}). One of Fermi's most
important results, to my view is the discovery that the highest energy
photons (in the LAT band) are observed to both (I) lag behind the
emission at lower energies; and (II) last longer. Both these results
are seen in Figure \ref{fig:080916c}. Similarly, the width of
individual pulses are energy dependent. It was found that the pulse
width $\omega$ vary with energy, $\omega(E) \propto E^{-\alpha}$ with
$\alpha \sim 0.3 - 0.4$ [\cite{Norris+05, Liang+06}].

Already in the BATSE era, several bursts were found to have
``ultra-long'' duration, having $T_{90}$ exceeding $\sim 10^3$~s
[e.g., \cite{Giblin+02, Nicastro+04}].  Recently, several additional
bursts were found in this category (e.g., GRB 091024A, GRB 101225A,
GRB 111209A GRB 121027A and GRB 130925A [\cite{Virgili+13, Gendre+13,
  Stratta+13, Levan+14, Evans+14}]), which raise the idea of a new class
of GRBs. If these bursts indeed represent a separate class, they may
have a different progenitor than that of ``regular'' long
GRBs [\cite{Gendre+13b, Nakauchi+13, Levan+14}]. However, recent analysis
showed that bursts with duration $T_{90} \sim 10^3$~s need not belong
to a special population, while bursts with $T_{90} \gsim 10^4$~s may
belong to a separate population [\cite{Zhang+14, GM14}]. As the
statistics is very low, my view is that this is still an open issue.

\subsection{Spectral properties}
\label{sec:spectra}

\subsubsection{A word of caution}

Since this is a rapidly evolving field, one has to be extra careful in
describing the spectra of GRB prompt emission. As I will show below,
the observed spectra is, in fact sensitive to the analysis method
chosen. Thus, before describing the spectra, one has to describe the
analysis method.

Typically, the spectral analysis is based on analyzing flux integrated
over the entire duration of the prompt emission, namely the spectra is
{\bf time-integrated}. Clearly, this is a trade off, as enough photons
need to be collected in order to analyze the spectra. For weak bursts
this is the only thing one can do. However, there is a major drawback
here: use of the time integrated spectra implies that important
time-dependent signals could potentially be lost or at least
smeared. This can easily lead to the wrong theoretical interpretation.

A second point of caution is the analysis method, which is done by a
forward folding technique. This means the following. First, a model
spectrum is chosen. Second, the chosen model is convolved with the
detector response, and compared to the detected counts
spectrum. Third, the model parameters are varied in search for the
minimal difference between model and data. The outcome is the best
fitted parameters within the framework of the chosen model.  This
analysis method is the only one that can be used, due to the
non-linearity of the detector's response matrix, which makes it
impossible to invert. 

However, the need to pre-determine the fitted model implies that the
results are biased by the initial hypothesis. {\textbf {Two different
    models can fit the data equally well.}} This fact, which is often
being ignored by theoreticians, is important to realize when the
spectral fits are interpreted. Key spectral properties such as the
energy of the spectral peak put strong constraints on possible
emission models. Below I show a few examples of different analysis
methods {\bf of the same data} that result in different spectral peak
energies, slopes, etc., and therefore lead to different theoretical
interpretations.

\subsubsection{The ``Band'' model}
\label{sec:Band}

In order to avoid biases towards a preferred physical emission model,
GRB spectra are traditionally fitted with a mathematical function,
which is known as the ``Band'' function (after the late David Band)
[\cite{Band+93}]. This function had become the standard in this field,
and is often refereed to as ``Band model''. The photon number spectra
in this model are given by:
\beq
N_{ph}(E) = A \left\{ \ba{ll}
\left( {E \over 100 \keV} \right)^\alpha \exp\left(-{E \over E_0}
\right) & E < (\alpha - \beta) E_0
\nonumber \\ 
\left[ {(\alpha -\beta) E_0 \over 100 \keV }\right]^{\alpha - \beta}
\exp\left(\beta - \alpha\right) \left({E \over 100 \keV} \right)^\beta
& E \geq (\alpha - \beta) E_0

\begin{figure}
\includegraphics[width=12cm]{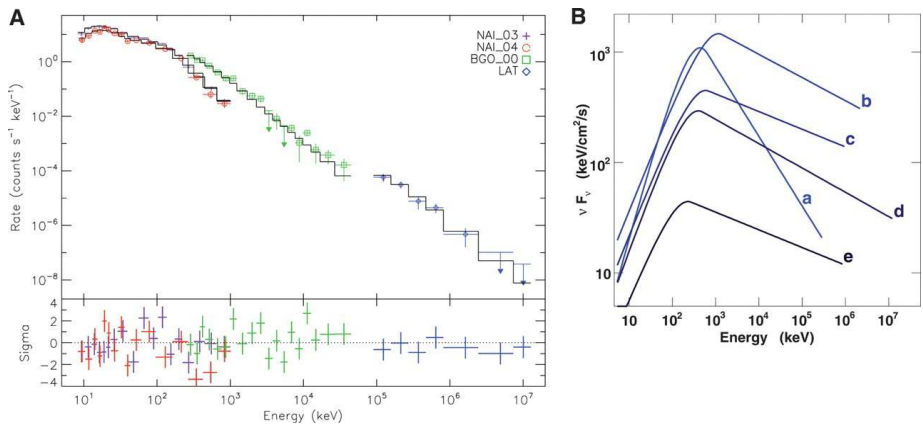}
\caption{ Spectra of GRB 080916C at the five time intervals (a-e)
  defined in Figure \ref{fig:080916c} are fitted with a ``Band''
  function. (A [left]): count spectrum for NaI, BGO, and LAT in time
  bin b. (B [right]) The model spectra in $\nu F_\nu$ units for all
  five time intervals, in which a flat spectrum would indicate equal
  energy per decade of photon energy, and the changing shapes show the
  evolution of the spectrum over time. The ``broken power law'' Figure
  adopted from \cite{Abdo+09b}.  }
\label{fig:080916c_spectra}
\end{figure}

\ea \right.  \eeq This model thus has 4 free parameters: low energy
spectral slope, $\alpha$, high energy spectral slope, $\beta$, break
energy, $\approx E_0$, and an over all normalization, $A$. It is found
that such a simplistic model, which resembles a ``broken power law''
is capable of providing good fits to many different GRB spectra; see
Figure \ref{fig:080916c_spectra} for an example. Thus, this model is
by far the most widely used in describing GRB spectra.

Some variations of this model have been introduced in the literature.
Examples are single power law (PL), ``smooth broken power law''
(SBPL), or ``Comptonized model'' (Comp) [see, e.g., \cite{Kaneko+06,
    Nava+11a, Goldstein+12, Goldstein+13}]. These are very similar in
nature, and do not, in general provide a better physical insight.

On the down side, clearly, having only 4 free parameters, this model
is unable to capture complex spectral behavior that is known now to
exist, such as the different temporal behavior of the high energy
emission discussed above. Even more importantly, as will be discussed
below, the limited number of free model parameters in this model can
easily lead to wrong conclusions. Furthermore, this model - on purpose
- is mathematical in nature, and therefore fitting the data with this
model does not, by itself, provide any clue about the physical origin
of the emission. In order to obtain such an insight, one has to
compare the fitted results to the predictions of different theoretical
models.

\begin{figure}
\includegraphics[width=4cm]{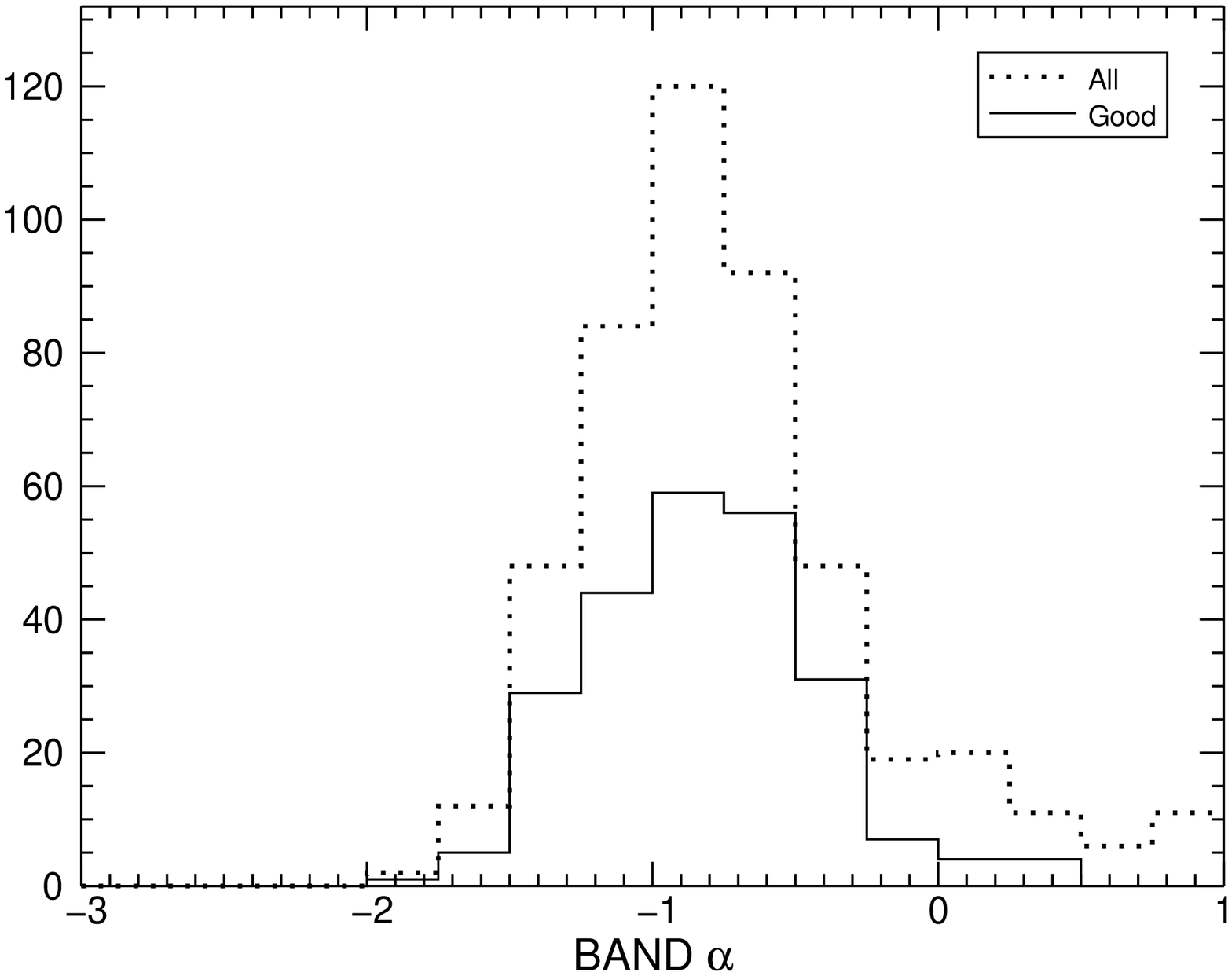}
\includegraphics[width=4cm]{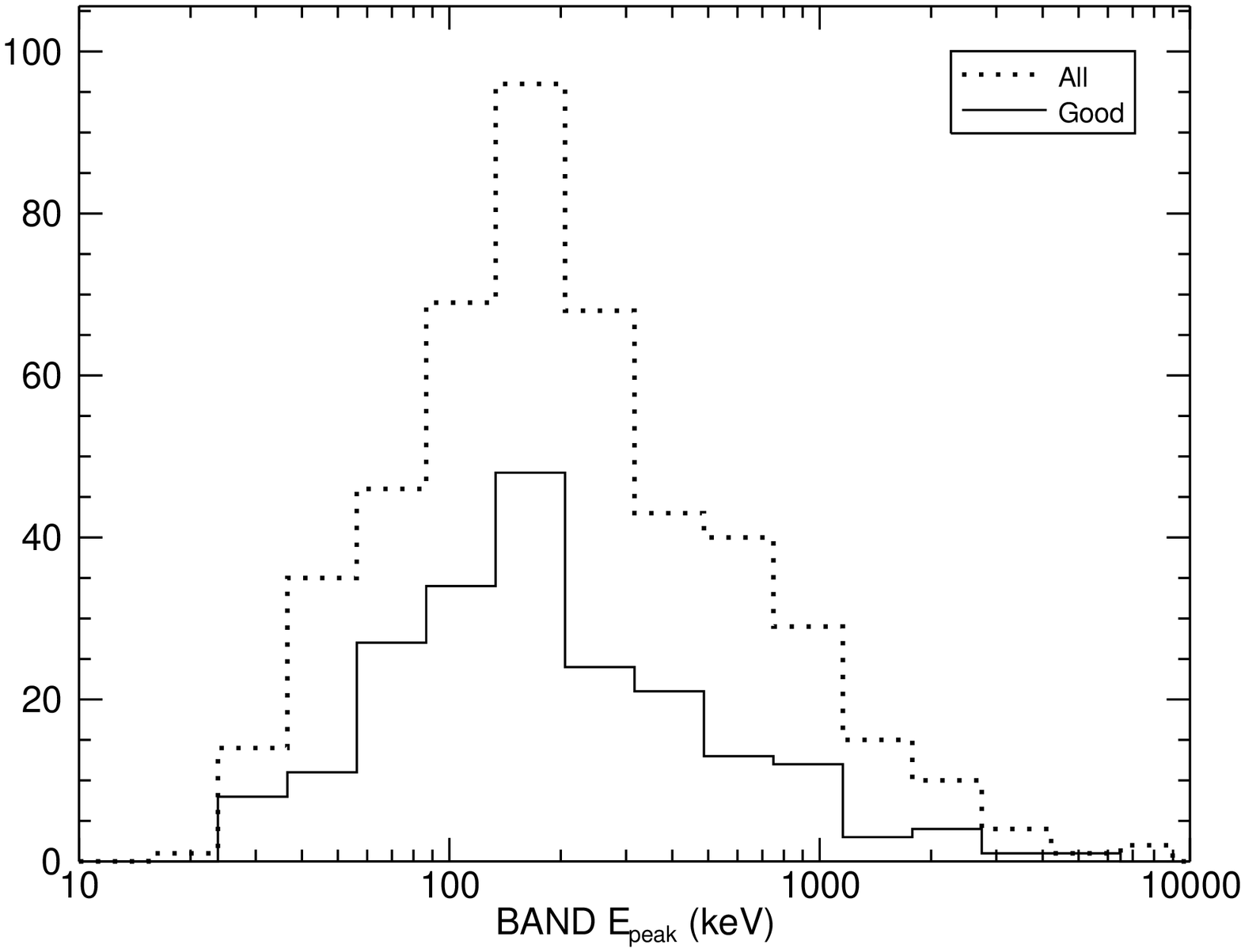}
\includegraphics[width=4cm]{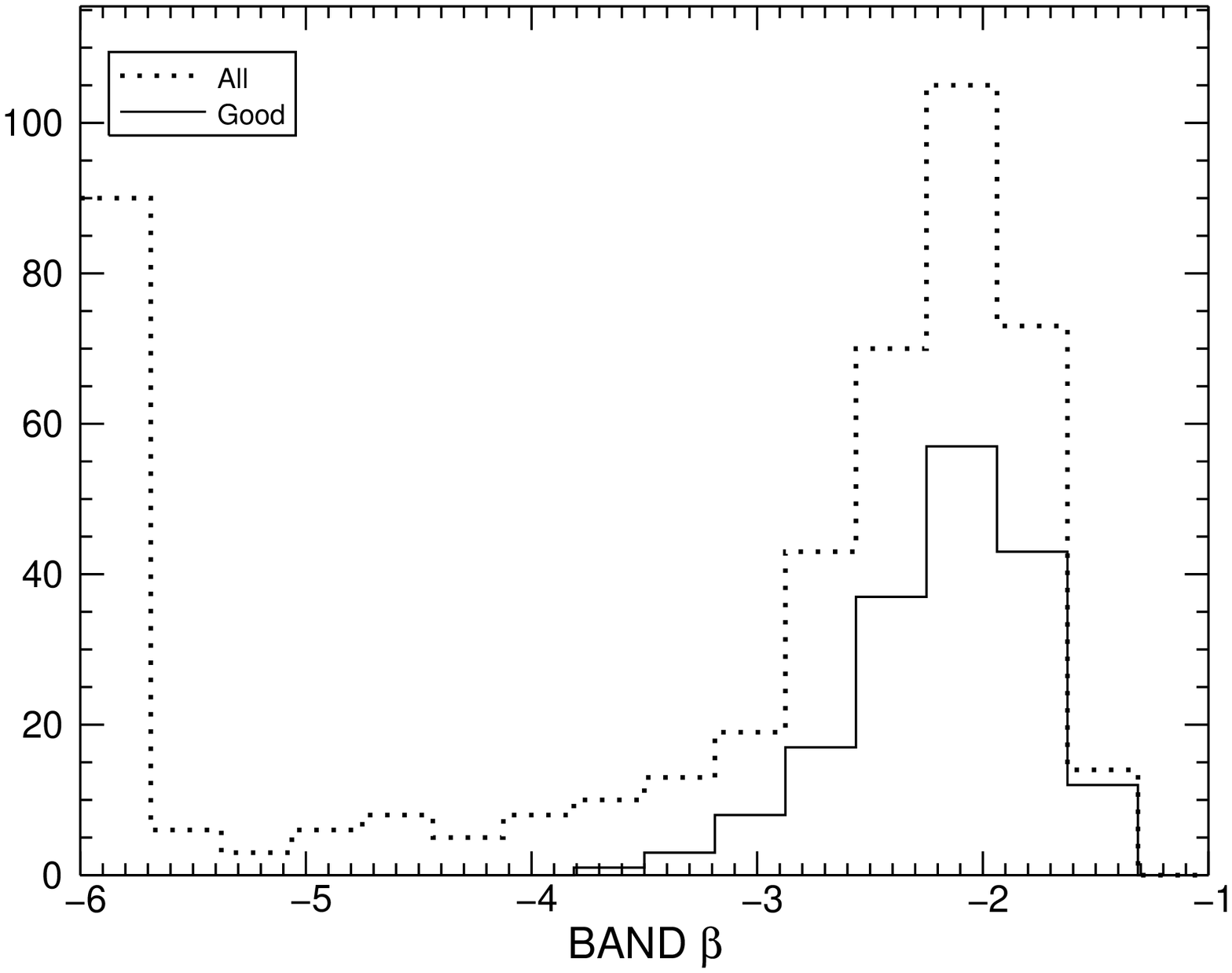}
\caption{Histograms of the distributions of ``Band'' model free
  parameters: the low energy slope $\alpha$ (left), peak energy
  $E_{peak}$ (center) and the high energy slope, $\beta$ (right). The
  data represent 3800 spectra derived from 487 GRBs in the first {\it
    FERMI}-GBM catalogue. The difference between solid and dashed
  curves are the goodness of fits- the solid curve represent fits
  which were done under minimum $\chi^2$ criteria, and the dash curves
  are for all GRBs in the catalogue. Figure adopted from
  \cite{Goldstein+12}.  }
\label{fig:BAND_distribution}
\end{figure}

When using the ``Band'' model to fit a large number of bursts, the
distribution of the key model parameters (the low and high energy
slopes $\alpha$ and $\beta$ and the peak energy $E_{peak}$) show a
surprisingly narrow distribution (see Figure
\ref{fig:BAND_distribution}). The spectral properties of the two
categories: short and long GRBs, detected by both {\it BATSE}, {\it
  Integral} as well as {\it Fermi} are very similar, with only minor
differences \citep{Preece+00, Kaneko+06, Nava+11, Zhang+11,
  Goldstein+12, Goldstein+13, Bosnjak+14, Gruber+14}. The low energy spectral
slope is roughly in the range $-1.5 < \alpha < 0$, averaging at
$\langle \alpha \rangle \simeq -1$. The distribution of the high
energy spectral slope peaks at $\langle \beta \rangle \simeq
-2$. While typically $\beta < -1.3$, many bursts show a very steep
$\beta$, consistent with an exponential cutoff. The peak energy
averages around $\langle E_{peak} \rangle \simeq 200$~keV, and it
ranges from tens keV up to $\sim$MeV (and even higher, in a few rare,
exceptional bursts).

As can be seen in Figures \ref{fig:080916c_spectra} and
\ref{fig:BAND_distribution}, the ``Band'' fits to the spectra have
three key spectral properties. (1) The prompt emission extends to very
high energies, $\gsim$~MeV. This energy is above the threshold for
pair production ($m_e = $~0.511 MeV), which is the original motivation
for relativistic expansion of GRB outflows (see below). (2) The
``Band'' fits do not resemble a ``Planck'' function; hence the reason
why thermal emission, which was initially suggested as the origin of
GRB prompt spectra [\cite{Goodman86, Pac86}] was quickly abandoned,
and not considered as a valid radiation process for a long time. (3)
The values of the free ``Band'' model parameters, and in particular
the value of the low energy spectral slope, $\alpha$ are not easily
fitted with any simply broad-band radiative process such as
synchrotron or synchrotron self-Compton (SSC). Although in some
bursts, synchrotron emission could be used to fit the spectra [e.g.,
  \cite{Tavani96, Cohen+97, PSM99, Frontera+00}], this is not the case
in the vast majority of GRBs [\cite{Crider+97, Preece+98, Preece+02,
    GCG03}].  This was noted already in 1998, with the term
``synchrotron line of death'' coined by R. Preece [\cite{Preece+98}],
to emphasis the inability of the synchrotron emission model to provide
good fits to the spectra of [most] GRBs.

Indeed, these three observational properties introduce a major
theoretical challenge, as currently no simple physically
motivated model is able to provide convincing explanation to the
observed spectra. However, as already discussed above, the ``Band''
fits suffer from several inherent major drawbacks, and therefore the
obtained results must be treated with great care.

\subsubsection{``Hybrid'' model}

\begin{figure}
\includegraphics[width=8cm]{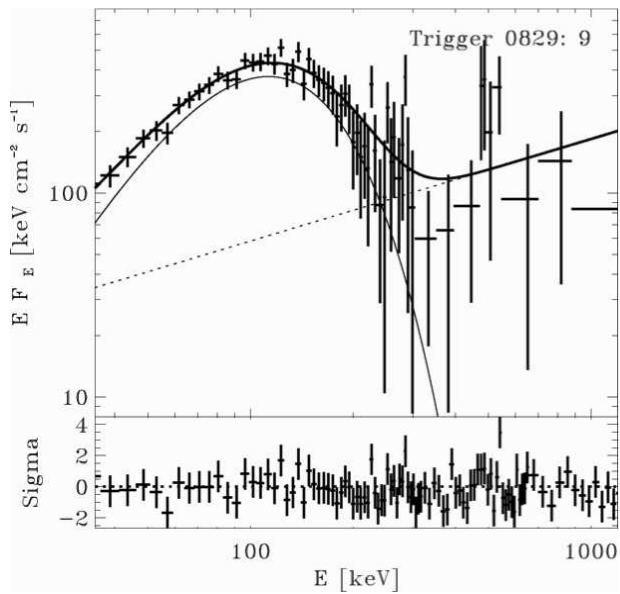}
\caption{A ``hybrid'' model fit to the spectra of GRB 910927 detected
  by BATSE.  Figure courtesy of F. Ryde.  }
\label{fig:hybrid_model}
\end{figure}

An alternative model for fitting the GRB prompt spectra was proposed
by F. Ryde [\cite{Ryde04, Ryde05}].  Being aware of the limitations of
the ``Band'' model, when analyzing BATSE data, Ryde proposed a
``hybrid'' model that contains a thermal component (a Planck function)
and a single power law to fit the non-thermal part of the spectra
(presumably, resulting from Comptonization of the thermal
photons). Ryde's hybrid model thus contain four free parameters - the
same number of free parameters as the ``Band'' model: two parameters
fit the thermal part of the spectrum (temperature and thermal flux)
and two fit the non-thermal part. Thus, as opposed to the ``Band''
model which is mathematical in nature, Ryde's model suggests a
physical interpretation to at least part of the observed spectra (the
thermal part). An example of the fit is shown in Figure
\ref{fig:hybrid_model}.

Clearly, a single power law cannot be considered a valid physical
model in describing the non-thermal part of the spectra, as it
diverges. Nonetheless, it can be acceptable approximation when
considering a limited energy range, as was available when analyzing
BATSE data. While the hybrid model was able to provide comparable, or
even better fits with respect to the ``Band'' model to several doesens
bright GRBs [\cite{Ryde04, Ryde05, RP09, McGlynn+09, Larsson+11}], is
was shown that this model over predict the flux at low energies (X-ray
range) for many GRBs [\cite{Ghirlanda+07, Frontera+13}]. This
discrepancy, however, can easily be explained by the
over-simplification of the use of a single power law as a way to
describe the non-thermal spectra both above and below the thermal
peak. From a physical perspective, one expects Comptonization to
modify the spectra above the thermal peak, but not below it; see
discussion below.

\begin{figure}
\includegraphics[width=8cm]{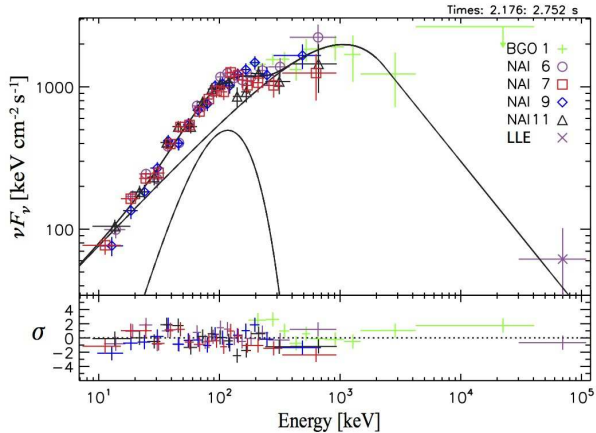}
\caption{The spectra of GRB110721A is best fit with a ``Band'' model (peaking
at $E_{peak} \sim 1$~MeV), and a blackbody component (having
temperature $T \sim 100$~keV). The advantage over using just a
``Band'' function is evident when looking at the residuals (Taken from
\cite{Iyyani+13}) }
\label{fig:Iyyani}
\end{figure}

As Fermi enables a much broader spectral coverage than BATSE, in
recent years Ryde's hybrid model could be confronted with data over a
broader spectral range. Indeed, it was found that in several bursts
(e.g., GRB090510 [\cite{Ackermann+10}], GRB090902B [\cite{Abdo+09a,
    Ryde+10, Ryde+11}] GRB110721A [\cite{Axelsson+12, Iyyani+13}],
GRB100724B [\cite{Guiriec+11}], GRB100507 [\cite{Ghirlanda+13}] or
GRB120323A [\cite{Guiriec+13}]) the broad band spectra are best fitted
with a combined ``Band + thermal'' model (see Figure
\ref{fig:Iyyani}). In these fits, the peak of the thermal component is
always found to be below the peak energy of the ``Band'' part of the
spectrum. This is consistent with the rising ``single power law'' that
was used in fitting the band-limited non thermal spectra.

The ``Band + thermal'' model fits require six free parameters, as
opposed to the four free parameters in both the ``Band'' and in the
original ``hybrid'' models. While this is considered as a drawback,
this model has several notable advantages. First, this model does not
suffer from the energy divergence of a single power law fit, as in
Ryde's original proposal. Second, in comparison with ``Band'' model
fits, it shows significant improvement in quality, both in statistical
errors (reduced $\chi^2$), and even more importantly, by the behavior
of the residuals: when fitting the data with a ``Band'' function,
often the residuals to the fit show a ``wiggly'' behavior, implying
that they are not randomly distributed. This is solved when adding the
thermal component to the fits.

Similar to Ryde's original model, fits with ``Band + thermal'' model
can provide a physical explanation to only the thermal part of the
spectra; they still do not suggest physical origin to the non-thermal
part of the spectra. Nonetheless, the addition of the thermal part
implies that the values of the free model parameters used in fitting
the non-thermal part, such as the low energy spectral slope
($\alpha$), as well as the peak energy $E_{peak}$ are different than
the values that would have been obtained by pure ``Band'' fits
(namely, without the thermal component; see \cite{Guiriec+13, BR14,
  DZ14, Guiriec+15}). In some bursts, the new values obtained are
consistent with the predictions of synchrotron theory, suggesting a
synchrotron origin of the non-thermal part [\cite{Burgess+14,
    Yu+15}]. However, in many cases this interpretation is
insufficient [e.g., \cite{BRY14}]; see further discussion
below. Another (relatively minor) drawback of these fits is that from
a theoretical perspective, even if a thermal component exists in the
spectra, it is expected to have the shape of a gray-body rather than a
pure ``Planck'', due to light aberration (see below).

One therefore concludes that the ``Band + thermal'' fits which became
very popular recently can be viewed as an intermediate step towards
full physically-motivated fits of the spectra. They contain a mix of a
physically-motivated part (the thermal part) with an addition
mathematical function (the ``Band'' part) whose physical origin still
needs clarification.

\begin{figure}
\includegraphics[width=24cm]{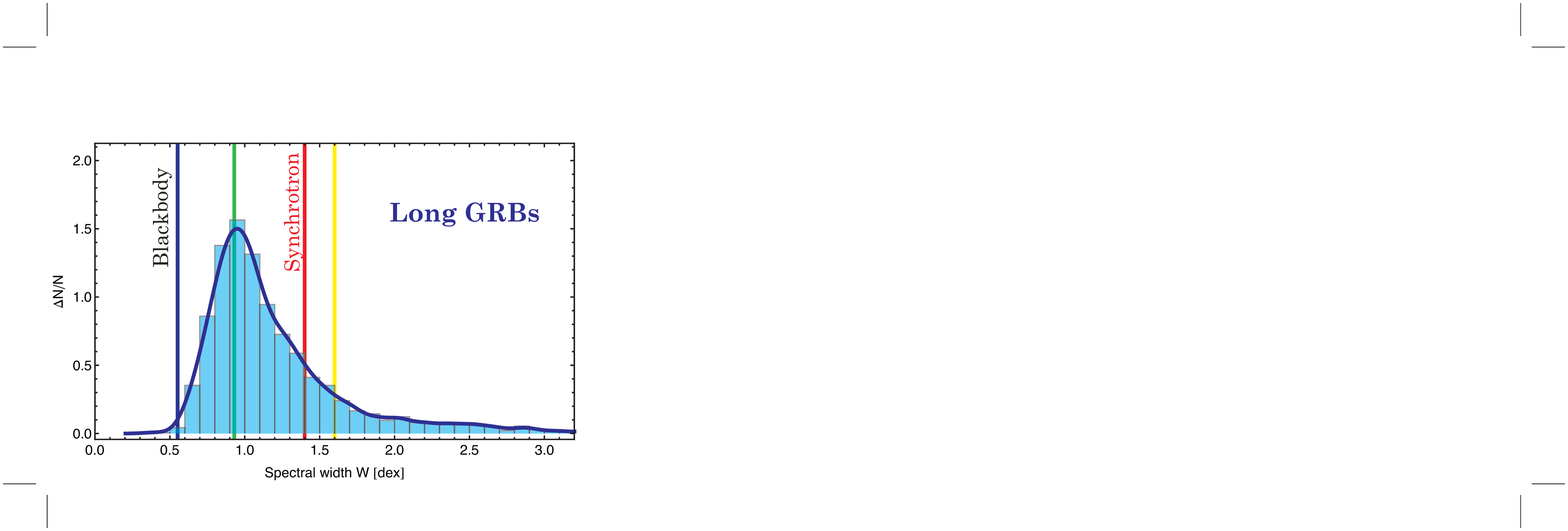}
\caption{Full width half maximum of the spectral peaks of over 2900
  bursts fitted with the ``Band'' function. The narrow most spectra
  are compatible with a ``Planck'' spectrum. About $\sim 80\%$ of the
  spectra are too narrow to be fitted with the (slow cooling)
  synchrotron emission model (red line). When fast cooling is added,
  nearly 100\% of the spectra are too narrow to be compatible with
  this model. As it is physically impossible to narrow the broad-band
  synchrotron spectra, these results thus suggest that the spectral
  peak is due to some widening mechanism on of the Planck spectrum,
  which are therefore pronounced (indirectly) in the vast majority of
  spectra. Figure taken from \cite{AB15}. }
\label{fig:Magnus}
\end{figure}

As of today, pure ``Planck'' spectral component is clearly identified
in only a very small fraction of bursts. Nonetheless, there is a good
reason to believe that it is in fact very ubiquitous, and that the
main reason it is not clearly identified is due to its distortion. A
recent work [\cite{AB15}], examined the width of the
spectral peak, quantified by $W$, the ratio of energies that define
the full width half maximum (FWHM). The results of an analysis of over
2900 different BATSE and Fermi bursts are shown in Figure
\ref{fig:Magnus}. The smaller $W$ is, the narrow the spectral
width. Imposed on the sample are the line representing the spectral
width from a pure ``Planck'' (black), and a line representing the
spectral width for slow cooling synchrotron (red). Fast cooling
synchrotron results in much wider spectral width, which would be shown
to the far right of this plot. Thus, while virtually all the spectral
width are wider than ``Planck'', over $\sim 80\%$ are narrower than
allowed by the synchrotron model. On the one hand, ``narrowing'' a
synchrotron spectra is (nearly) impossible. However, there are various
ways, which will be discussed below in which pure ``Planck'' spectra
can be broadened. Thus, although ``pure'' Planck is very rare, these
data suggests that broadening of the ``Planck'' spectra plays a major
role in shaping the spectral shape of the vast majority of GRB
spectra.

\subsubsection {Time resolved spectral analysis}
 
Ryde's original analysis is based on time-resolved spectra. The
lightcurve is cut into time bins (having typical duration $\gsim
1$~s), and the spectra at each time bin is analyzed
independently. This approach clearly limits the number of bursts that
could be analyzed in this method to only the brightest ones,
presumably those showing smooth lightcurve over several - several tens
of seconds (namely, mainly the long GRBs). However, its great
advantage is that it enables to detect temporal evolution in the
properties of the fitted parameters; in particular, in the temperature
and flux of the thermal component.

\begin{figure}
\includegraphics[width=6cm]{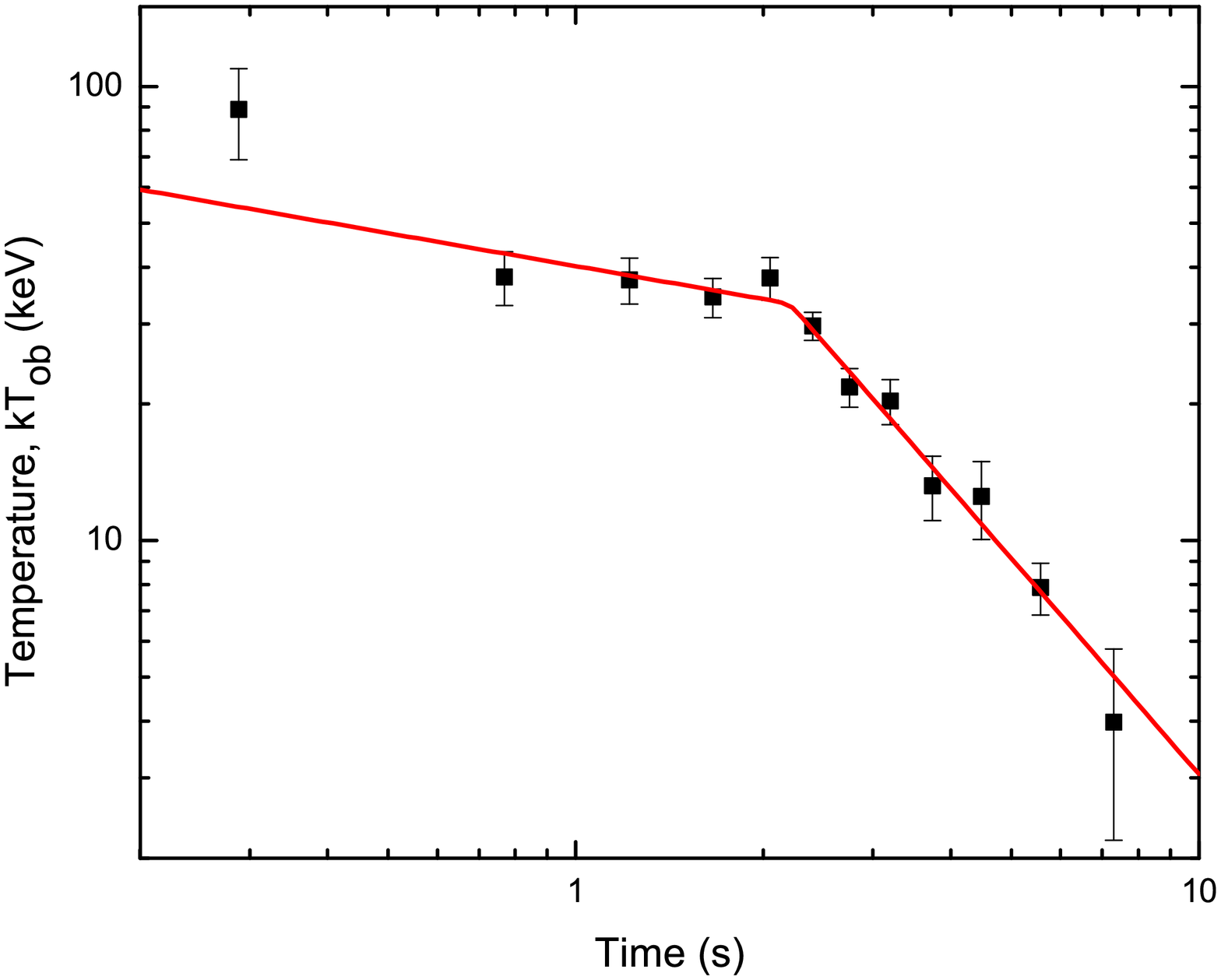}
\includegraphics[width=6cm]{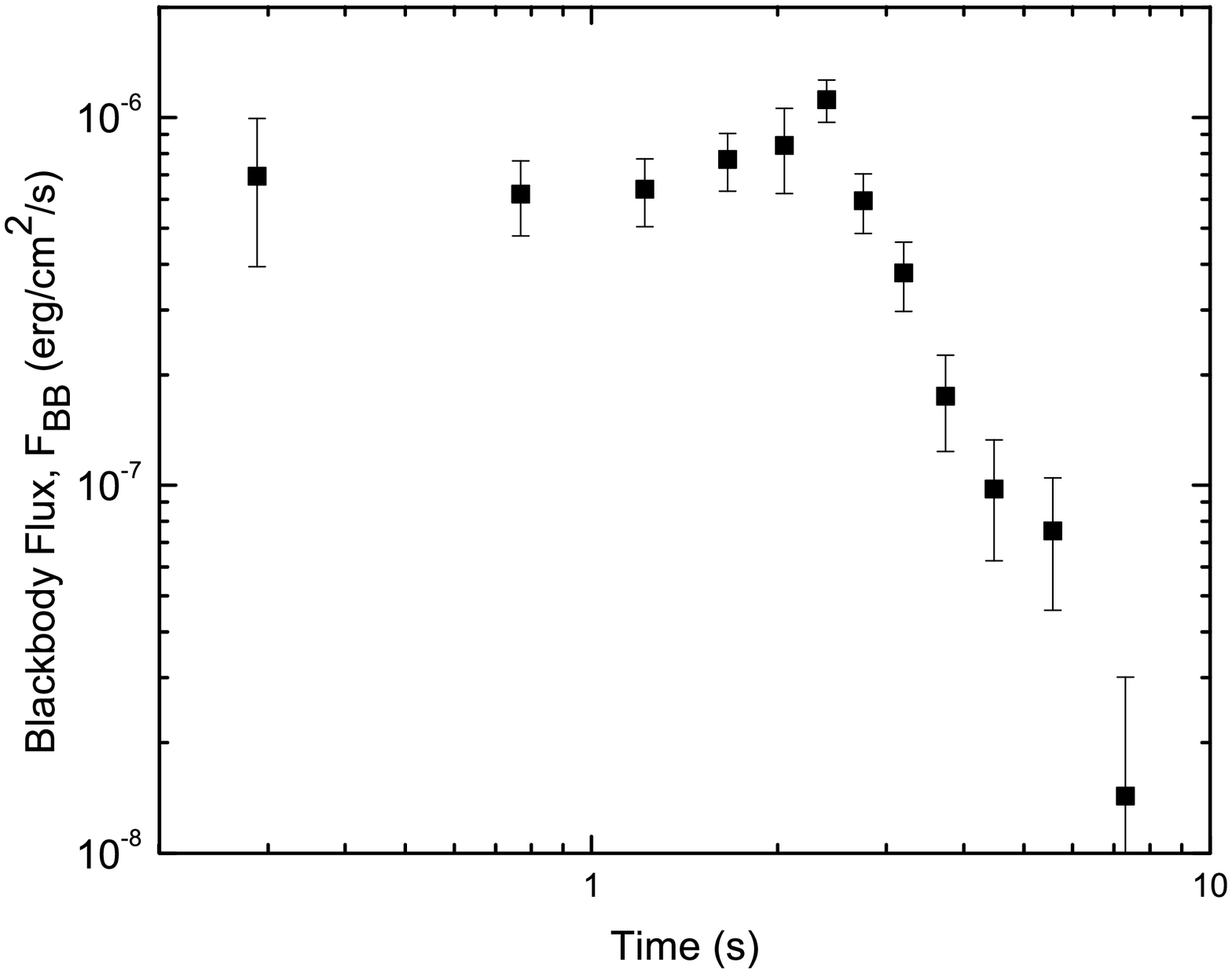}
\caption{{\bf Left:} the temperature of the thermal component of
  GRB110721A at different time bins show a clear ``broken power law''
  with $T(t) \sim t^{-0.25}$ before $t_{brk} \sim 3$~s, and $T(t) \sim
  t^{-0.67}$ at later times. {\bf Right:} the flux of the thermal
  component show a similar broken power law temporal behavior, with
  similar break time. At late times, $F_{BB}(t) \propto t^{-2}$. See
  \cite{RP09} for details. Figure courtesy of F. Ryde. }
\label{fig:T_t}
\end{figure}

One of the key results of the analysis carried by \cite{RP09}, is the
well defined temporal behavior of both the temperature and flux of the
thermal component. Both the temperature and flux evolve as a broken
power law in time: $T\propto t^\alpha$, and $F \propto t^\beta$, with
$\alpha \simeq 0$ and $\beta \simeq 0.6$ at $t < t_{brk} \approx $few
s, and $\alpha \simeq -0.68$ and $\beta \simeq -2$ at later times (see
Figure \ref{fig:T_t}). This temporal behavior was found among all
sources in which thermal emission could be identified. It may
therefore provide a strong clue about the nature of the prompt
emission, in at least those GRBs for which thermal component was
identified. To my personal view, these findings may hold the key to
understanding the origin of the prompt emission, and possibly the
nature of the progenitor.

Due to Fermi's much greater sensitivity, time resolved spectral
analysis is today in broad use. This enables to observed temporal
evolution not only of the thermal component, but of other parts of the
spectra as well (see, e.g., Figure \ref{fig:080916c_spectra}).  As an
example, a recent analysis of GRB130427A reveals a temporal change in
the peak energy during the first 2.5~s of the burst, which could be
interpreted as due to synchrotron origin [\cite{Preece+14}].

\subsubsection{Distinguished high energy component}
\label{sec:high_energy_component}

\begin{figure}
\includegraphics[width=8cm]{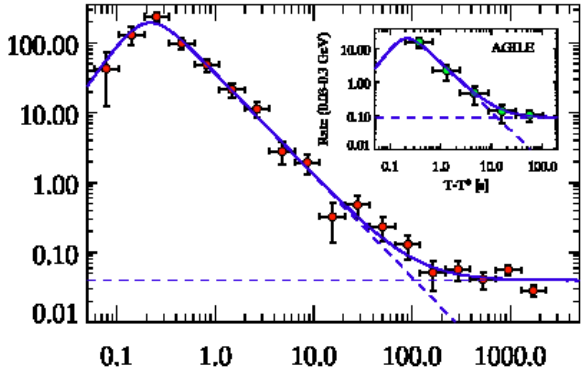}
\caption{Lightcurve of the emission of GRB090510 above 100 \MeV
  extends to $> 100$~s, and can be fitted with a smoothly broken power
  law. The times are scaled to the time $T^\star = 0.6$~s after the
  GBM trigger. The inset shows the AGILE lightcurve (energy range
  30-300 \MeV), extending to much shorter times. Figure taken from
  \cite{Ghirlanda+10}. }
\label{fig:090510}
\end{figure}

Prior to the {\it Fermi} era, time resolved spectral analysis was very
difficult to conduct due to the relatively low sensitivity of the {\it
  BATSE} detector, and therefore its use was limited to bright GRBs
with smooth lightcurve. However, {\it Fermi's} superb sensitivity
enables to carry time resolved analysis to many more bursts.
 One of the findings is the delayed onset of GeV
emission with respect to emission at lower energies which is seen in a
substantial fraction of LAT bursts (see, e.g.,
Figure \ref{fig:080916c}). This delayed onset is further accompanied by
a long lived emission ($\gsim 10^2$~s), and separate lightcurve
[\cite{Abdo+09a, Abdo+09b, Abdo+09c, KB10}]. The GeV emission decays as
a power law in time, $L_{\rm GeV} \propto t^{-1.2}$ [\cite{Ghirlanda+10,
    Ackerman+13, Nava+14}]. Furthermore, the GeV emission shows smooth
decay (see Figure \ref{fig:090510}). This behavior naturally points
towards a separate origin of the GeV and lower energy photons; see
discussion below.

Thus, one can conclude that at this point in time (Dec. 2014), evidence
exist for three separate components in GRB spectra: (I) a thermal
component, peaking typically at $\sim 100$~keV; (II) a non-thermal
component, whose origin is not fully clear, peaking at $\lsim $~MeV
and - lacking clear physical picture, is fitted with a ``Band''
function; and (III) a third component, at very high energies ($\gsim
100$~MeV) showing a separate temporal evolution [\cite{Zhang+11, Guiriec+15}]. 

Not all three components are clearly identified in all GRBs; in fact,
separate evolution of the high energy part is observed in only a
handful of GRBs. The fraction of GRBs which show clear evidence for
the existence of a thermal component is not fully clear; it seem to
depend on the brightness, with bright GRBs more likely to show
evidence for a thermal component (up to 50\% of bright GRBs show clear
evidence for a separate thermal component [\cite{Guiriec+15}
and Larsson et. al., in prep.]). Furthermore, this fraction is
sensitive to to the analysis method. Thus, final conclusions are still
lacking.

Even more interestingly, it is not at all clear that the ``bump''
identified as a thermal component is indeed such; such a bump could
have other origins as well (see discussion below). Thus, I think it is
fair to claim that we are now in a transition phase: on the one hand,
it is clear that fitting the data with a pure ``Band'' model is
insufficient, and thus more complicated models, which are capable of
capturing more subtle features of the spectra are being used. On the
other hand, these models are still not fully physically motivated, and
thus a full physical insight of the origin of prompt emission is still
lacking.

\subsection{Polarization}
\label{sec:polarization}

The leading models of the non-thermal emission, namely synchrotron
emission and Compton scattering, both produce highly polarized
emission [\cite{RL79}]. Nonetheless, due to the spherical assumption,
the inability to spatially resolve the sources, and the fact that
polarization was initially discovered only during the afterglow phase
[\cite{Covino+99, Wijers+99}], polarization was initially discussed
only in the context of GRB afterglow, but not the prompt phase [e.g.,
  \cite{LP98, GW99, GL99, ML99, GrK03}].
  
The first claim of highly linearly polarized prompt
emission in a GRB, $\Pi = (80 \pm 20)\%$ in GRB021206 by RHESSI
[\cite{CB03}] was disputed by a later analysis [\cite{RF04}]. A later
analysis of BATSE data show that the prompt emission of GRB930131 and
GRB96092 are consistent with having high linear polarization, $\Pi >
35\% $ and $\Pi > 50\%$; though the exact degree of polarization could
not be well constrained [\cite{Willis+05}]. Similarly, \cite{Kalemci+07,
  McGlynn+07} and \cite{Gotz+09} showed that the prompt spectrum of
GRB041219a observed by INTEGRAL is consistent with being highly
polarized, but with no statistical significance.

Recently, high linear polarization, $\Pi = (27 \pm 11)\%$ was observed
in the prompt phase of GRB 100826a by the GAP instrument on board
IKAROS satellite [\cite{Yonetoku+11}].  As opposed to former
measurements, the significance level of this measurement is high, $2.9
\sigma$. High linear polarization degree was further detected in
GRB110301a ($\Pi = 70 \pm 22 \% $) with $3.7 \sigma$ confidence, and
in GRB100826a ($\Pi = 84^{+16}{}_{-28}\%$) with $3.3 \sigma$
confidence [\cite{Yonetoku+12}].

As of today, there is no agreed theoretical interpretation to the
observed spectra (see discussion below). However, different
theoretical models predict different levels of polarization, which are
correlated with the different spectra. Therefore, polarization
measurements have a tremendous potential in shedding new light on the
different theoretical models, and may hold the key in discriminating
between them.

\subsection{Emission at other wavebands}
\label{sec:wavebands}

Clearly, the prompt emission spectra is not necessarily limited to
those wavebands that can be detected by existing satellites. Although
broad band spectral coverage is important in providing clues to the
origin of the prompt emission and the nature of GRBs, due to their
random nature and to the short duration it is extremely difficult to
observe the prompt emission without fast, accurate triggering.

As the physical origin of the prompt emission is not fully clear, it
is difficult to estimate the flux at wavebands other than
observed. Naively, the flux is estimated by interpolating the ``Band''
function to the required energy [e.g., \cite{Granot+10}].  However, as
discussed above (and proved in the past), this method is misleading,
as (1) the ``Band'' model is a very crude approximation to a more
complicated spectra; and (2) the values of the ``Band'' model low and
high energy slopes change when new components are added. Thus, it is of
no surprise that early estimates were not matched by observations.

\subsubsection{High energy counterpart}

At high energies, there has been one claim of possible TeV emission
associated with GRB970417a [\cite{Atkins+00}]. However, since then, no
other confirmed detection of high energy photons associated with any
GRB prompt emission were reported. Despite numerous attempts, only
upper limits on the very high energy flux were obtained by the
different detectors (MAGIC: [\cite{Albert+07a, Aleksic+14}], MILAGRO:
[\cite{Milagro+07}], HESS: [\cite{Aharonian+09a, Aharonian+09b,
    Abramowski+14}], VERITAS: [\cite{Acciari+11}], HAWC:
[\cite{Abeysekara+15}]).

\subsubsection{Optical counterpart} 

At lower energies (optic -- X), there have been several long GRBs for
which a precursor (or a very long prompt emission duration) enabled
fast slew of ground based robotic telescopes (and / or Swift XRT and
UVOT detectors) to the source during the prompt phase. The first ever
detection of optical emission during the prompt phase of a GRB was
that of GRB990123 [\cite{Akerlof+99}]. Other examples of optical
detection are GRB041219A [\cite{Blake+05}], GRB060124
[\cite{Romano+06}], GRB 061121 [\cite{Page+07}] the ``naked eye'' GRB080319B
[\cite{Racusin+08}] GRB080603A [\cite{Guidorzi+11}] GRB080928
[\cite{Rossi+11}] GRB090727 [\cite{Kopac+13}] GRB121217a
[\cite{Elliott+14}] GRB1304a7A [\cite{Maselli+14}] GRB130925a
[\cite{Greiner+14}] for a partial list.

\begin{figure}
\includegraphics[width=6cm]{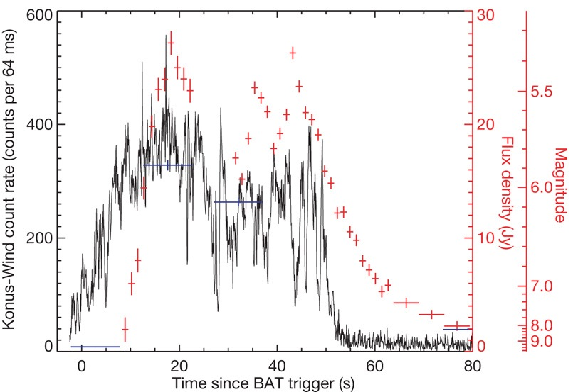}
\includegraphics[width=6cm]{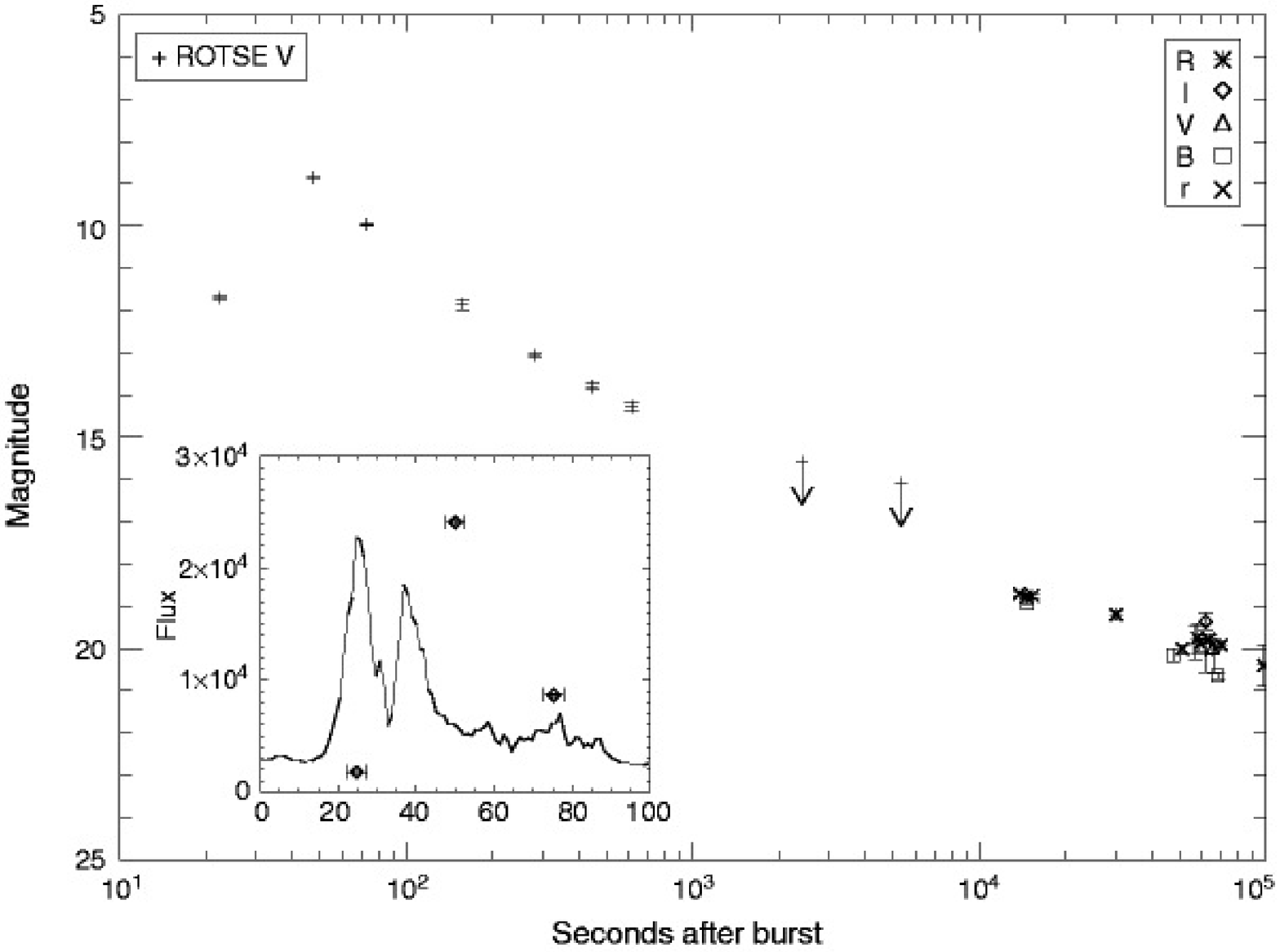}
\caption{Left: Gamma-ray lightcurve (black) and optical data from 'Pi
  in the sky' (blue) and 'Tortora' (red) of GRB080319B, show how the
  optical component traces the $\gamma$-ray component. Figure taken
  from \cite{Racusin+08}. Right: Gamma-ray and optical lightcurve of
  GRB990123 show that the optical lightcurve lags behind the
  $\gamma$-rays. Figure taken from \cite{Akerlof+99}.  }
\label{fig:080319B}
\end{figure}

\begin{figure}
\includegraphics[width=6cm]{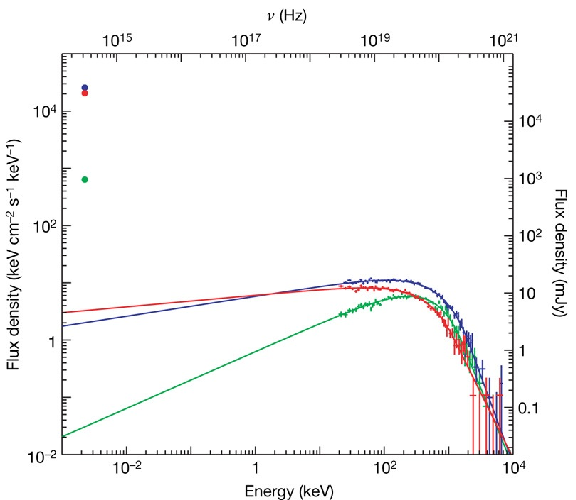}
\includegraphics[width=6cm]{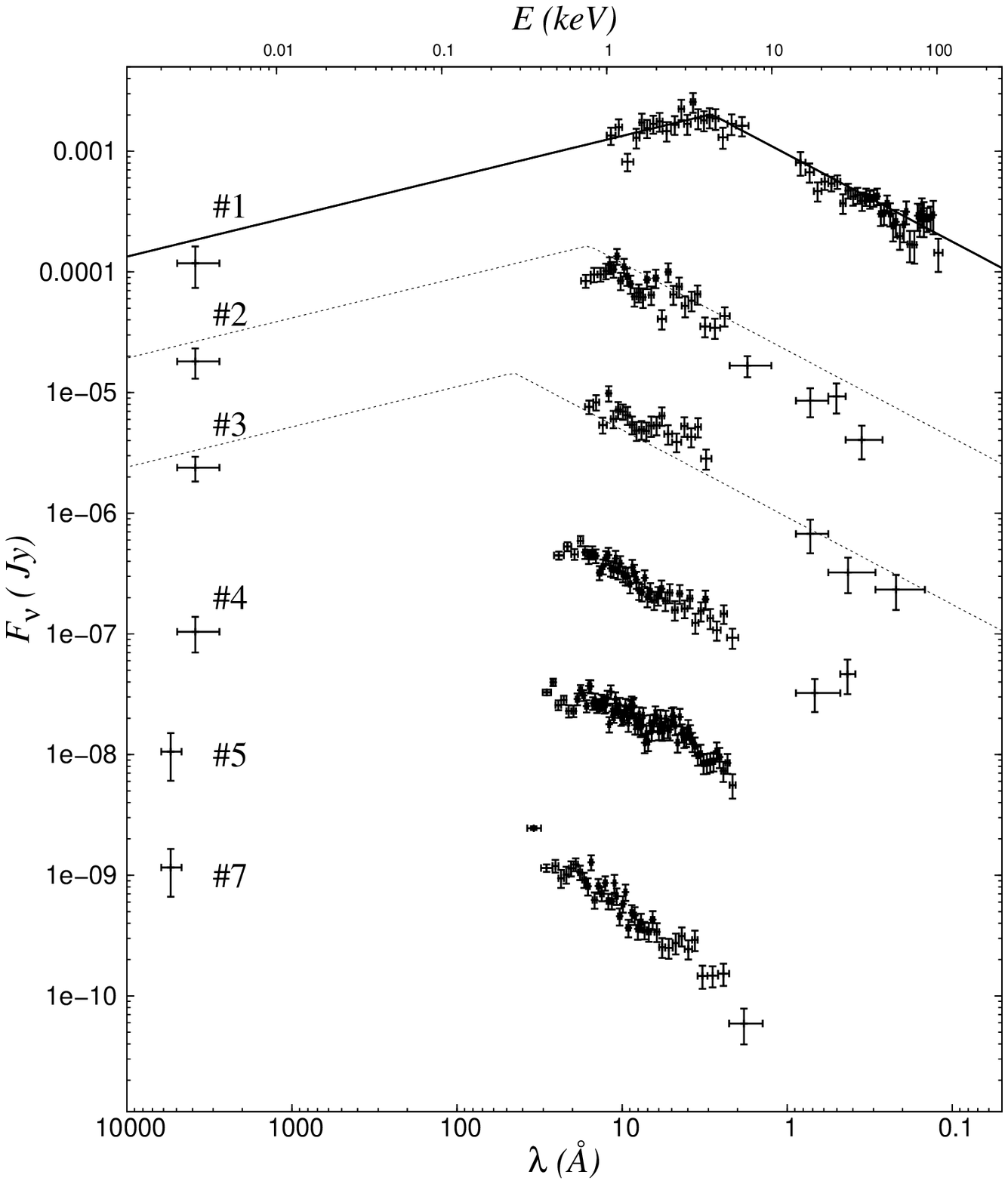}
\caption{Left: the 'Pi in the Sky' and Konus-wind flux at 3 time
  intervals (fitted by a ``Band'' model) of GRB080319B. The optical
  flux is 2-3 orders of magnitude above the direct
  interpolation. Figure taken from \cite{Racusin+08}. Right: combined
  UVOT and X/$\gamma$ ray data of GRB080928 at early times are fitted
  with a broken power law. For this burst, the slope is consistent
  with having synchrotron origin. Figure taken from \cite{Rossi+11}.
}
\label{fig:080319B_spectra}
\end{figure}

The results are diverse. In some cases (e.g., GRB990123), the peak of
the optical flux lags behind that of the $\gamma$-ray flux, while in
other GRBs (e.g., GRB080319B), no lag is observed. This is shown in
Figure \ref{fig:080319B}. Similarly, while in some bursts, such as
GRB080319B or GRB090727 the optical flux is several orders of
magnitude higher than that obtained by direct interpolation of the
``Band'' function from the $x/\gamma$ ray band, in other bursts, such
as GRB080928, it seem to be fitted well with a broken-power law
extending at all energies (see Figure \ref{fig:080319B_spectra}).  To
further add to the confusion, some GRBs show complex temporal and
spectral behavior, in which the optical flux and lightcurve changes
its properties (with respect to the x/$\gamma$) emission with
time. Examples are GRB050820 [\cite{Vestrand+06}] and GRB110205A
[\cite{Zheng+12}].

These different properties hint towards different origin of the
optical emission. It should be stressed that due to the observational
constraints, optical counterparts are observed to date only in very
long GRBs, with typical $T_{90}$ of hundreds of seconds (or
more). Thus, the optical emission may be viewed as part of the prompt
phase, but also as part of the early afterglow; it may result from the
reverse shock which takes place during the early afterglow epoch. See
further discussion below.

\subsection{Correlations}
\label{sec:correlations}

There have been several claims in the past for correlations between
various observables of the prompt GRB emission. Clearly, such
correlations could potentially be extremely useful in both
understanding the origin of the emission, as well as the ability to
use GRBs as probes, e.g., ``standard candles'' similar to supernova
1a. However, a word of caution is needed: as already discussed, many
of the correlations are based on values of fitted parameters, such as
$E_{pk}$, which are sensitive to the fitted model chosen - typically,
the ``Band'' function. As more refined models - such as, e.g., the
addition of a thermal component can change the peak energy, the
claimed correlation may need to be modified.  Since final conclusion
about the best physically motivated model that can describe the prompt
emission spectra has not emerged yet, it is too early to know the
modification that may be required to the claimed
correlations. Similarly, some of the correlations are based on the
prompt emission duration, which is ill-defined.

The first correlation was found between the peak energy (identified as
temperature) and luminosity of single pulses within the prompt
emission [\cite{Golenetskii+83}]. They found $L \propto
E_{peak}^\alpha$, with $\alpha \sim 1.6$. These results were confirmed
by \cite{Kargatis+94}, though the errors on $\alpha$ were large, as
$\alpha \simeq 1.5 - 1.7$.

A similar correlation between the (redshift corrected) peak energy and
the (isotropic equivalent) total gamma-ray energy of different bursts
was reported by \cite{Amati+02}, namely $E_{peak,z} \propto
E_{\gamma,iso}^\alpha$, with $\alpha \sim 0.5$ [\cite{Amati+02,
    Amati06, Frontera+12}]. Here, $E_{peak,z} = E_{peak} (1+z)$. This
became known as the ``Amati relation''.

The Amati relation has been questioned by several authors, claiming
that it is an artifact of a selection effect or biases [e.g.,
  \cite{NP05, BP05, Butler+07, Butler+09, SN11, Collazzi+11,
    Kocevski12}]. However, counter arguments are that even is such
selection effects exist, they cannot completely exclude the
correlation [\cite{Ghirlanda+05, Ghirlanda+08, Ghirlanda+12, Nava+12b,
    BR13}]. To conclude, it seem that current data (and analysis
method) do support some correlation, though with wide scatter. This
scatter still needs to be understood before the correlation could be
used as a tool, e.g., for cosmological studies [\cite{Virgili+12,
    HAZ13}].

There are a few other notable correlations that were found in recent
years. One is a correlation between the (redshift corrected) peak
energy $E_{peak,z}$ and the isotropic luminosity in $\gamma$-rays at
the peak flux, $L_{\gamma,peak,iso}$ [\cite{WG03, Yonetoku+04}]:
$E_{peak,z} \propto L_{\gamma,p,iso}^{0.52}$.  A second correlation is
between $E_{peak,z}$ and the geometrically-corrected gamma-ray energy,
$E_\gamma \simeq (\theta_j^2/2) E_{\gamma, iso}$, where $\theta_j$ is
the jet opening angle (inferred from afterglow observations):
$E_{peak,z} \propto E_\gamma^{0.7}$ [\cite{Ghirlanda+04}]. It was
argued that this relation is tighter than the Amati relation; however,
it relies on the correct interpretation of breaks in the afterglow
lightcurve to be associated with jet breaks, which can be problematic
[\cite{Ghisellini+07, Liang+08, KB08, Racusin+09, Ryan+15}].

Several other proposed correlations exist; I refer the reader to
\cite{KZ14}, for a full list.

\section{Theoretical framework}
\label{sec:theory}

\begin{figure}
\includegraphics[width=11cm]{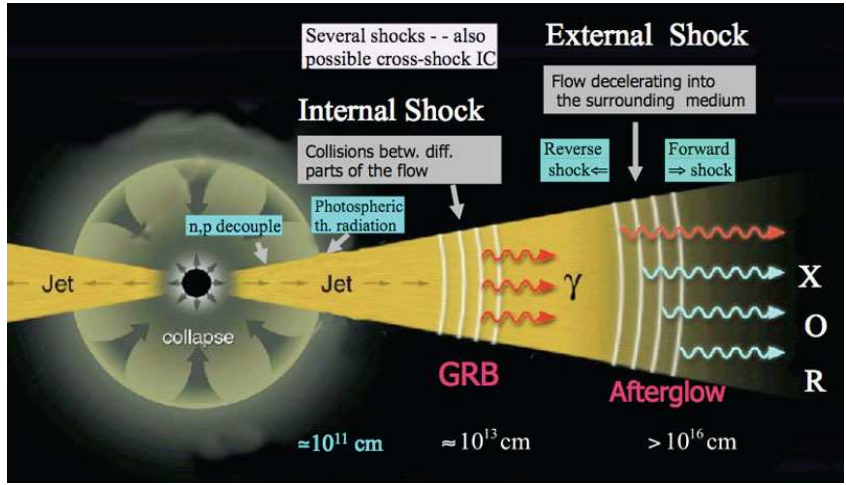}
\caption{ Cartoon showing the basic ingredients of the GRB
  ``fireball'' model. (1) The source of energy is a collapse of a
  massive star (or merger of NS-NS or NS-BH, not shown here). (2) Part
  of this energy is used in producing the relativistic jet. This could
  be mediated by hot photons (``fireball''), or by magnetic field. (3)
  The thermal photons decouple at the photosphere. (4) Part of the jet
  kinetic energy is dissipated (by internal collisions, in this
  picture) to produce the observed $\gamma$ rays. (5) The remaining
  kinetic energy is deposited into the surrounding medium, heating it
  and producing the observed afterglow.  Cartoon is taken from
  \cite{MR14}.  }
\label{fig:fireball}
\end{figure}

Perhaps the easiest way to understand the nature of GRBs is to follow
the different episodes of energy conversion. Although the details of
the energy transfer are still highly debatable, there is a wide
agreement, based on firm observational evidence, that there are
several key episodes of energy conversion in GRBs.  (1) Initially, a
large amount of energy, $\sim 10^{53}$~erg or more, is released in a
very short time, in a compact region. The source of this energy must
be gravitational. (2) Substantial part of this energy is converted
into kinetic energy, in the form of relativistic outflow. This is the
stage in which GRB jets are formed and accelerated to relativistic
velocities. The exact nature of this acceleration process, and in
particular the role played by magnetic fields in it, is still not
fully clear. (3) (Part of) this kinetic energy is dissipated, and is
used in producing the gamma rays that we observe in the prompt
emission. Note that part of the observed prompt emission (the thermal
part) may originate directly from photons emitted during the initial
explosion; the energy carried by these photons is therefore not
initially converted to kinetic form.  (4) The remaining of the kinetic
energy (still in the form of relativistic jet) runs into the
interstellar medium (ISM) and heats it, producing the observed
afterglow. The kinetic energy is thus gradually converted into heat,
and the afterglow gradually fades away. A cartoon showing these basic
ingredients in the context of the ``fireball'' model, is shown in
Figure \ref{fig:fireball}, adapted from \cite{MR14}.

\subsection{Progenitors}
\label{sec:progenitors}

The key properties that are required from GRB progenitors are: (1)
the ability to release a huge amount of energy, $\sim
10^{52}-10^{53}$~erg (possibly even larger), within the observed GRB
duration of few seconds; (2) the ability to explain the fast time
variability observed, $\delta t \gsim 10^{-3}$~s, implying (via light
crossing time argument) that the energy source must be compact: $R
\sim c \delta t \sim 300$~km, namely of stellar size.

While 20 years ago, over hundred different models were proposed in
explaining possible GRB progenitors [see \cite{Nemiroff94}], natural
selection (namely, confrontation with observations over the years) led
to the survival of two main scenarios. The first is a merger of two
neutron stars (NS-NS), or a black hole and a neutron star (BH-NS). The
occurance rate, as well as the expected energy released, $\sim G M^2/R
\sim 10^{53}$~erg (using $M \sim M_\odot$ and $R \gsim R_{sch.}$, the
Scharzschild radius of stellar-size black hole), are sufficient for
extra-galactic GRBs [\cite{Eichler+89, Pac90, Narayan+91, MR92,
    Narayan+92}]. The alternative scenario is the core collapse of a
massive star, accompanied by accretion into a black hole
[\cite{Woosley93, Pac98, Pac98b, Fryer+99, MW99, Popham+99, WB06} and
  references therein]. In this scenario, similar amount of energy, up
to $\sim 10^{54}$~erg may be released by tapping the rotational energy
of a Kerr black hole formed in the core collapse, and/or the inner
layers of the accretion disk.

The observational association of long GRBs to type Ib/c supernova
discussed above, as well as the time scale of the collapse event,
$\lsim 1$~minute, which is similar to that observed in long GRBs, makes
the core collapse, or ``collapsar'' model, the leading model for
explaining long GRBs. The merger scenario, on the other hand, is
currently the leading model in explaining short GRBs [see, e.g.,
  discussions in \cite{Nakar07, GRF09, Berger14}].

\subsection{Relativistic expansion and kinetic energy dissipation: the ``fireball'' model}
\label{sec:expansion}

A GRB event is associated with a catastrophic energy release of a
stellar size object. The huge amount of energy, $\sim 10^{52}-
10^{53}$~erg released in such a short time and compact volume, results
in a copious production of neutrinos - antineutrinos (initially in
thermal equilibrium) and possible release of gravitational
waves. These two, by far the most dominant energy forms are of yet not
detected. A smaller fraction of the energy (of the order $10^{-3} -
10^{-2}$ of the total energy released) goes into high temperature
($\gsim$~MeV) plasma, containing photons, $e^\pm$ pairs, and baryons,
known as ``fireball'' [\cite{CR78}]. The fireball may contain a
comparable - or even larger amount of magnetic energy, in which case
it is Poynting flux dominated [\cite{Usov94, Thompson94, Katz97,
    MR97b, LB03, ZY11}] \footnote{some authors use the phrase ``cold
  fireball'' in describing magnetically-dominated ejecta, as opposed
  to ``hot fireballs''; here, I will simply use the term ``fireball''
  regardless of the fraction of energy stored in the magnetic field.}.

The scaling laws that govern the expansion of the fireball depend on
its magnetization. Thus, one must discriminate between
photon-dominated (or magnetically-poor) outflow and magnetic dominated
outflow. I discuss in this section the photon-dominated (``hot
fireball''). Magnetic dominated (``cold fireball'') will be discussed
in the next section (section \ref{sec:magnetized_flow}).

\subsubsection{Photon dominated outflow}

Let us consider first photon-dominated outflow. In this model, it is
assumed that a large fraction of the energy released during the
collapse / merger is converted directly into photons close to the jet
core, at radius $r_0$ (which should be $\gsim$ the Schwarzschild radius
of the newly formed black hole).  The photon temperature is
\beq
T_0 = \left( L \over 4 \pi r_0^2 c a \right)^{1/4} = 1.2 \,
L_{52}^{1/4} r_{0,7}^{-1/2} \MeV
\label{eq:T0}
\eeq
where $a$ is the radiation constant, $L$ is the luminosity and $Q =
10^x Q_x$ in cgs units is used here and below. This temperature is
above the threshold for pair production, implying that a large number
of $e^\pm$ pairs are created via photon-photon interactions (and
justifying the assumption of full thermalization).  The observed
luminosity is many orders of magnitude above the Eddington luminosity,
$L_E = 4 \pi G M m_p c/ \sigma_T = 1.25 \times 10^{38} (M/M_\odot)
{\rm ~erg\, s^{-1}}$, implying that radiation pressure is much larger
than self gravity, and the fireball must expand.

The dynamics of the expected relativistic fireball were first
investigated by [\cite{Goodman86, Pac86, SP90}]. The ultimate velocity
it will reach depends on the amount of baryons (baryon load) within
the fireball [\cite{Pac90}], which is uncertain. The baryon load can
be deduced from observations: as the final expansion kinetic energy
cannot exceed the explosion energy, the highest Lorentz factor that
can be reached is $\Gamma_{\max} = E/M c^2$. Thus, the fact that GRBs are
known to have high bulk Lorentz factors, $\Gamma \gsim 10^2$ at later
stages (during the prompt and afterglow emission) [\cite{KP91,
    Fenimore+93, WL95, BH97, SP99, LS01, ZKM03, Molinari+07, Liang+10,
    Racusin+11}] imply that only a small fraction of the baryons in
the progenitor star(s) are in fact accelerated and reach relativistic
velocities.

\subsubsection{Scaling laws for relativistic expansion: instantaneous energy release}

The scaling laws for the fireball evolution follows conservation of
energy and entropy.  Let us assume first that the energy release is
``instantaneous'', namely within a shell of size $\delta r \sim
r_0$. Thus, the total energy contained within the shell (as seen by an
observer outside the expanding shell) is
\beq
E^{ob} \propto \Gamma(r) V' T'(r)^4 \propto \Gamma(r) r^2 \Gamma(r)
r_0 T'(r)^4 = const.
\label{eq:energy}
\eeq
Here, $T'(r)$ is the shell's comoving temperature, and $V' = 4 \pi r^2
\delta r'$ is its comoving volume.  Note that the first factor of
$\Gamma(r)$ is needed in converting the comoving energy to the
observed energy, and the second originates from transformation of the
shell's width: the shell's comoving width (as measured by a comoving
observer within it) is related to its width as measured in the lab
frame ($r_0$) by $\delta r' = \Gamma(r) r_0$.

Starting from the fundamental thermodynamic relation, $dS = (dU + p
dV)/T$, one can write the entropy of a fluid component with zero
chemical potential (such as photon fluid) in its comoving frame, $S' = V'(u'
+p')/T'$. Here, $u'$, $p'$ are the internal energy density and
pressure measured in the comoving frame. For photons, $p'=u'/3 \propto
T'^4$. Since initially, both the rest mass and energy of the baryons
are negligible, the entropy is provided by the photons. Thus,
conservation of entropy implies
\beq
S' \propto V' T'(r)^3 \propto r^2 \Gamma(r) r_0 T'(r)^3 = const. 
\label{eq:entropy}
\eeq
Dividing Equations \ref{eq:energy} and \ref{eq:entropy}, one obtains
$\Gamma(r) T'(r) = const$, from which (using again these equations)
one can write the scaling laws of the fireball evolution,
\beq
T'(r) \propto r^{-1}~;~~ \Gamma(r) \propto r~;~~V'(r) \propto r^3
\eeq  

As the shell accelerates, the baryon kinetic energy $\Gamma(r) M c^2$
increases, until it become comparable to the total fireball energy
(the energy released in the explosion) at $\Gamma = \Gamma_{\max}
\simeq \eta$, at radius $r_s \sim \eta r_0$ (assuming that the outflow
is still optically thick at $r_s$, and so the acceleration can
continue until this radius). Here, $\eta \equiv E/M c^2$ is the
specific entropy per baryon. Note that during the acceleration phase,
the shell's kinetic energy increase comes at the expense of the
(comoving) internal energy, as is reflected by the fact that the
comoving temperature drops.

Beyond the saturation radius $r_s$, most of the available energy is in
kinetic form, and so the flow can no longer accelerate, and it
coasts. The spatial evolution of the Lorentz factor is thus
\beq
\Gamma(r) = \left\{
\ba{ll}
(r/r_0) & ~~~r\lsim r_s; \\
\eta   & ~~~r \gsim r_s. 
\ea
\right.
\eeq

Equation \ref{eq:entropy} that describes conservation of (comoving)
entropy, holds in this regime as well; therefore, in the regime $r >
r_s$ one obtains $r^2 T'(r)^3 = const$, or
\beq
\Gamma(r) = \eta ~;~~ T'(r) \propto r^{-2/3} ~;~~V'(r) \propto r^2.
\eeq
The observed temperature therefore evolves with radius as 
\beq
T^{ob}(r) = \Gamma(r) T'(r) =  \left\{
\ba{ll}
T_0 &~~~ r < r_s; \\
T_0 \times (r/r_s)^{-2/3} &~~~ r > r_s
\ea
\right.
\label{eq:8}
\eeq

\subsubsection{Continuous energy release}
\label{sec:scalings2}

Let us assume next that the energy is released over a longer duration,
$t \gg r_0/c$ (as is the case in long GRBs). In this scenario, the
progenitor continuously emits energy at a rate $L$ (erg/s), and this
emission is accompanied by mass ejected at a rate $\dot M = L/\eta
c^2$. The analysis carried above is valid for each fluid element
separately, provided that $E$ is replaced by $L$ and $M$ by $\dot M$,
and thus the scaling laws derived above for the evolution of the
(average) Lorentz factor and temperature as a function of radius
hold. However, there are a few additions to this scenario.

We first note the following [\cite{Waxman03}]. The comoving number
density of baryons follow mass conservation:
\beq
n'_p(r) = {\dot M \over 4 \pi r^2 m_p c \Gamma(r)} = {L \over 4 \pi
  r^2 m_p c^3 \eta \Gamma(r)}
\label{eq:n}
\eeq
(assuming spherical explosion).
Below $r_s$, the (comoving) energy density of each fluid element is
relativistic, $aT'(r)^4/n_p' m_p c^2 = \eta (r_0/r)$. Thus, the speed of
sound in the comoving frame is $c_s \simeq c/\sqrt{3} \sim c$. The
time it takes a fluid element to expand to radius $r$, $r/c$ in the
observer frame, corresponds to time $t' \sim r/\Gamma c$ in the
comoving frame; during this time, sound waves propagate a distance $t'
c_s \sim r c / \Gamma c = r / \Gamma$ (in the comoving frame), which
is equal to $r/ \Gamma^2 = r_0^2 / r$ in the observer frame. This
implies that at the early stages of the expansion, where $r \gsim
r_0$, sound waves have enough time to smooth spatial fluctuations on
scale $\sim r_0$. On the other hand, regions separated by $\Delta r >
r_0$ cannot interact with each other. As a result, fluctuations in the
energy emission rate would result in the ejection and propagation of a
collection of independent sub-shells, each have typical thickness
$r_0$.

Each fluid element may have a slightly different density, and thus
have a slightly different terminal Lorentz factor; the standard
assumption is $\delta \Gamma \sim \eta$. This implies a velocity
spread $\delta v = v_1 - v_2 \approx {c \over 2 \eta^2}$, where $\eta$
is the characteristic value of the terminal Lorentz factor.  If such
two fluid elements originate within a shell (of initial thickness
$r_0$), spreading between these fluid elements will occur after
typical time $t_{spread} = r_0/\delta v$, and at radius (in the
observer's frame) [\cite{MLR93}]
\beq
r_{spread} = v_2 t_{spread} \simeq c r_0 \left({2 \eta^2
\over c }\right) \simeq 2 \eta^2 r_0
\eeq
According to the discussion above, this is also the typical radius
where two separate shells will begin to interact (sometimes referred
to as the ``collision radius'', $r_{col}$).

The spreading radius is a factor $\eta$ larger than the
saturation radius.  Thus, no internal collisions are expected during
the acceleration phase, namely at $r<r_s$. Below the spreading radius
individual shell's thickness (in the observer's frame), $\delta r$, is
approximately constant and equal to $r_0$.  At larger radii, $r>
r_{spread}$, it becomes $\delta r = r \delta v/c \sim r/\eta^2$.

Since the comoving radial width of each shell is $\delta r' = \Gamma
\delta r$, it can be written as
\beq
\delta r' \sim \left\{ \ba{ll}
r_0 \Gamma \sim r & ~~~ r< r_s \\
r_0 \eta & ~~~ r_s < r < r_{spread} \\
r/\eta  & ~~~ r > r_{spread}
\ea
\right.
\eeq
The comoving volume of each sub-shell, $V' \propto r^2 \delta r'$ is thus
 \beq
V' \propto r^2 \delta r' \sim \left\{ \ba{ll}
r^3 & ~~~ r< r_s \\
r_0 \eta r^2 & ~~~ r_s < r < r_{spread} \\
r^3/\eta  & ~~~ r > r_{spread}
\ea
\right.
\eeq

\subsubsection{Internal collisions as possible mechanism of kinetic energy dissipation}

At radii $r > r_{spread} = r_{coll}$, spreading within a single shell,
as well as interaction between two consecutive shells become
possible. The idea of shell collision was suggested early on
[\cite{PX94, RM94, SP97, Kobayashi+97, DM98}], as a way to dissipate the jet
  kinetic energy, and convert it into the observed radiation.

The key advantages of the internal collision model are: (1) its
simplicity - it is a very straight forward idea that naturally rises
from the discussion above; (2) it is capable of explaining the rapid
variability observed; and (3) the internal collisions are accompanied
by (internal) shock waves. It is believed that these shock waves are
capable of accelerating particles to high energies, via Fermi
mechanism. The energetic particles, in turn, can emit the high-energy,
non-thermal photons observed, e.g., via synchrotron emission. Thus,
the internal collisions is believed to be an essential part in this
energy conversion chain that results in the production of
$\gamma$-rays.

On the other hand, the main drawbacks of the model are (1) the very
low efficiency of energy conversion; (2) by itself, the model does not
explain the observed spectra - only suggests a way in which the
kinetic energy can be dissipated. In order to explain the observed
spectra, one needs to add further assumptions about how the dissipated
energy is used in producing the photons (e.g., assumptions about
particle acceleration, etc.). Furthermore, as will be discussed in
section \ref{sec:radiation} below, it is impossible to explain the observed spectra
within the framework of this model using standard radiative processes
(such as synchrotron emission or Compton scattering), without invoking
additional assumptions external to it. (3) Another major drawback of
this model is its lack of predictivity: while it does suggest a way of
dissipating the kinetic energy, it does not provide many details,
such as the time in which dissipations are expected, or the amount
of energy that should be dissipated in each collision (only rough
limits). Thus, it lacks a predictive power.

The basic assumption is that at radius $r_{coll} = r_{spread}$ two
shells collide. This collision dissipates part of the kinetic energy,
and converts it into photons. The time delay of the produced photons
(with respect to a hypothetical photon emitted at the center of
expansion and travels directly towards the observer) is
\beq
\delta t^{ob} \simeq {r_{coll} \over 2 \eta^2 c} \sim {r_0 \over c},
\eeq 
namely is of the same order as the central engine variability
time. Thus, this model is capable of explaining the rapid ($\gsim
1$~ms) variability observed.

On the other hand, this mechanism suffers a severe efficiency problem,
as only the differential kinetic energy between two shells can be
dissipated. Consider, e.g., two shells of masses $m_1$ and $m_2$, and
initial Lorentz factors $\Gamma_1$ and $\Gamma_2$ undergoing plastic
collision.  Conservation of energy and momentum implies that the final
Lorentz factor of the combined shell is [\cite{Kobayashi+97}]
\beq
\Gamma_f \simeq \left( {m_1 \Gamma_1 + m_2 \Gamma_2 \over {m_1 /
    \Gamma_1} + {m_2 / \Gamma_2} }\right)^{1/2} 
\eeq 
(assuming that both $\Gamma_1, \Gamma_2 \gg 1$).

The efficiency of kinetic energy dissipation is 
\beq
\eta = 1 - {(m_1 + m_2) \Gamma_f \over m_1 \Gamma_1 + m_2 \Gamma_2}
\simeq 1 - {m_1 + m_2 \over \left( m_1^2 + m_2^2 + m_1 m_2
  \left({\Gamma_1\over \Gamma_2} + {\Gamma_2 \over \Gamma_1}\right)
  \right)^{1/2} }
\eeq
Thus, in order to achieve high dissipation efficiency, one ideally
requires similar masses, $m_1 \simeq m_2$ and high contrast in Lorentz
factors $(\Gamma_1/\Gamma_2) \gg 1$. Such high contrast is difficult
to explain within the context of either the ``collapsar'' or the
``merger'' progenitor scenarios.

Even under these ideal conditions, the combined shell's Lorentz
factor, $\Gamma_f$ will be high; therefore the contrast between the
Lorentz factors of a newly coming shell and the merger shell in the
next collision, will not be as high. As a numerical example, if the
initial contrast is $(\Gamma_1/\Gamma_2) =10$, for $m_1 =m_2$ one can
obtain high efficiency of $\gsim 40\%$; however, the efficiency of the
next collision will drop to $\sim 11\%$. When considering ensemble of
colliding shells under various assumptions of the ejection properties
of the different shells, typical values of the global efficiency are
of the order of $1\% - 10\%$.  [\cite{MMM95, Kobayashi+97, PSM99,
    LGC99, Kumar99, Spada+00, GSE01, MZ09}]

These values are in contrast to observational evidence of a much
higher efficiency of kinetic energy conversion during the prompt
emission, of the order of tens of percents ($\sim~50\%$), which are
inferred by estimating the kinetic energy using afterglow measurements
[\cite{LZ04, Ioka+06, Nousek+06, Zhang+07b, NFP09, Peer+12}].

While higher efficiency of energy conversion in internal shocks was
suggested by a few authors [\cite{Beloborodov00, KS01}], we point out
that these works assumed very large contrast in Lorentz factors,
$(\Gamma_1/\Gamma_2) \gg 10$ for almost all collisions; as discussed
above such a scenario is unlikely to be realistic within the framework
of the known progenitor models.

I further stress that the efficiency discussed in this section
refers only to the efficiency in dissipating the kinetic
energy. There are a few more episodes of energy conversion that are
required before the dissipated energy is radiated as the observed
$\gamma$-rays. These include (i) using the dissipated energy to
accelerate the radiating particles [likely electrons] to high
energies; (ii) converting the radiating particle's energy into
photons; and (iii) finally, the detectors are sensitive only over a
limited energy band, and thus part of the radiated photons cannot be
detected. Thus, over all, the measured efficiency, namely, the energy
of the observed $\gamma$-ray photons relative to the kinetic energy,
is expected to be very low in this model, inconsistent with observations.

An alternative idea for kinetic energy dissipation arises from the
possibility that the jet composition may contain a large number of free
neutrons. These neutrons, that are produced by dissociation of nuclei
by $\gamma$-ray photons in the inner regions, decouple from the
protons below the photosphere (see below) due to the lower cross
section for proton-neutron collision relative to Thomson cross section
[\cite{Derishev+99, BM00, MR00b, Rossi+06}].  This leads to friction
between protons and neutrons as they have different velocities, which,
in turn results in production of $e^+$ that follow the decay of pions
(which are produced themselves by $p-n$ interactions). These positrons
IC scatter the thermal photons, producing $\gamma$-ray radiation
peaking at $\sim$~MeV [\cite{Beloborodov10}]. A similar result is
obtained when non-zero magnetic fields are added, in which case
contribution of synchrotron emission becomes comparable to that of
scattering the thermal photons [\cite{Vurm+11}].

\subsubsection{Optical depth and photosphere}
\label{sec:photosphere}

During the initial stages of energy release, a high temperature,
$\gsim$~MeV (see Equation \ref{eq:T0}) ``fireball'' is formed. At such
high temperature, large number of $e^\pm$ pairs are produced
[\cite{Pac86, Goodman86, SP90}]. The photons are scattered by these
pairs, and cannot escape. However, once the temperature drops to $T'
\lsim 17$~keV, the pairs recombine, and thereafter only a residual
number of pairs is left in the plasma [\cite{Pac86}]. Provided that $\eta
\lsim 10^5$, the density of residual pairs is much less than the
density of ``baryonic'' electrons associated with the protons, $n_e =
n_p$. (A large number of pairs may be produced later on, when kinetic
energy is dissipated, e.g., by shell collisions). This recombination
typically happens at $r< r_s$.

Equation \ref{eq:n} thus provides a good approximation to the number
density of both protons and electrons in the plasma. Using this
equation, one can calculate the optical depth by integrating the mean
free path of photons emitted at radius $r$. A 1-d calculation (namely,
photons emitted on the line of sight) gives [\cite{Pac90, ANP91}]:
\beq
\tau = \int_r^\infty n_e' \sigma_T \Gamma (1- \beta) dr' \simeq n_e'
\sigma_T {r \over 2 \Gamma},
\eeq
where $\beta$ is the flow velocity, and $\sigma_T$ is Thomson's cross
section; the use of this cross section is justified since in the
comoving frame, the photon's temperature is $T' = T^{ob}/\Gamma \ll m_e c^2$.

The photospheric radius can be defined as the radius from which
$\tau(r_{ph}) = 1$,
\beq
r_{ph} \simeq {\dot M \sigma_T \over 8 \pi r^2 m_p c \Gamma \eta} = {L
  \sigma_T \over 8 \pi m_p c^3 \Gamma \eta^2} \simeq 2 \times
10^{11}~L_{52}\,\eta_{2.5}^{-3}~{\rm cm}.
\label{eq:r_ph}
\eeq
In this calculation, I assumed constant Lorentz factor $\Gamma =
\eta$, which is justified for $r_{ph} > r_s$. In the case of
fluctuative flow resulting in shells, $\eta$ represents an average
value of the shell's Lorentz factor. Further note that an upper
limit on $\eta$ within the framework of this model is given by the
requirement $r_{ph} > r_s \rightarrow \eta < (L \sigma_T/ 8 \pi m_p
c^3 r_0) \simeq 10^3~L_{52}^{1/4} r_{0,7}^{-1/4}$. This is because as
the photons decouple the plasma at the photosphere, for larger values
of $\eta$ the acceleration cannot continue above $r_{ph}$ [\cite{MR00,
    MRRZ02}]. In this scenario, the observed spectra is expected to be
(quasi)-thermal, in contrast to the observations.

The observed temperature at the photosphere is calculated using
Equations \ref{eq:T0}, \ref{eq:8} and \ref{eq:r_ph},
\beq
T^{ob} = T_0 \left({r_{ph} \over r_s}\right)^{-2/3} = {80 \over
  (1+z)} ~~L_{52}^{-5/12} \eta_{2.5}^{8/3} r_{0,7}^{1/6}~{\rm keV}.
\label{eq:T_ob}
\eeq
Similarly, the observed thermal luminosity, $L_{th}^{ob} \propto r^2
\Gamma^2 T'^4 \propto r^0$ at $r<r_s$ and $L_{th}^{ob} \propto
r^{-2/3}$ at $r>r_s$ [\cite{MR00}]. Thus,
\beq
{L_{th} \over L} \simeq \left({r_{ph} \over r_s}\right)^{-2/3} = 6.6
\times 10^{-2}\, L_{52}^{-2/3} \eta_{2.5}^{8/3} r_{0,7}^{2/3}.
\label{eq:L_th}
\eeq
Note the very strong dependence of the observed temperature and
luminosity\footnote{Here, $L$ is the luminosity released in the
  explosion; the observed luminosity in $\gamma$-rays is just a
  fraction of this luminosity.} on $\eta$.

The results of Equation \ref{eq:L_th} show that the energy released as
thermal photons may be a few \% of the explosion energy. This value is
of the same order as the efficiency of the dissipation of kinetic
energy via internal shocks. However, as discussed above, only a
fraction of the kinetic energy dissipated via internal shocks is
eventually observed as photons, while no additional episodes of energy
conversion (and losses) are added to the result in Equation
\ref{eq:L_th}. Furthermore, the result in Equation \ref{eq:L_th} is
very sensitive to the uncertain value of $\eta$, via the ratio of
$(r_{ph}/r_s)$: for high $\eta$, $r_{ph}$ is close to $r_s$, reducing
the adiabatic losses and increasing the ratio of thermal
luminosity. In such a scenario, the internal shocks - if occurring, are
likely to take place at $r_{coll} \sim \eta r_s > r_{ph}$, namely in
the optically thin region. I will discuss the consequences of this
result in section \ref{sec:phot_emission} below.

The calculation of the photospheric radius in Equation \ref{eq:r_ph}
was generalized by \cite{Peer08} to include photons emitted off-axis;
in this case, the term ``photospheric radius'' should be replaced with
``photospheric surface'', which is the surface of last scattering of
photons before they decouple the plasma. Somewhat counter intuitively,
for a relativistic ($\Gamma \gg 1$) spherical explosion this surface
assumes a parabolic shape, given by [\cite{Peer08}]
\beq
r_{ph}(\theta) \simeq {R_d \over 2 \pi} \left( {1 \over \Gamma^2} +
{\theta^2 \over 3} \right),
\label{eq:r_ph_theta}
\eeq
where $R_d \equiv {\dot M} \sigma_T / (4 m_p \beta c)$ depends on the
mass ejection rate and velocity. 

\begin{figure}
\includegraphics[width=10cm]{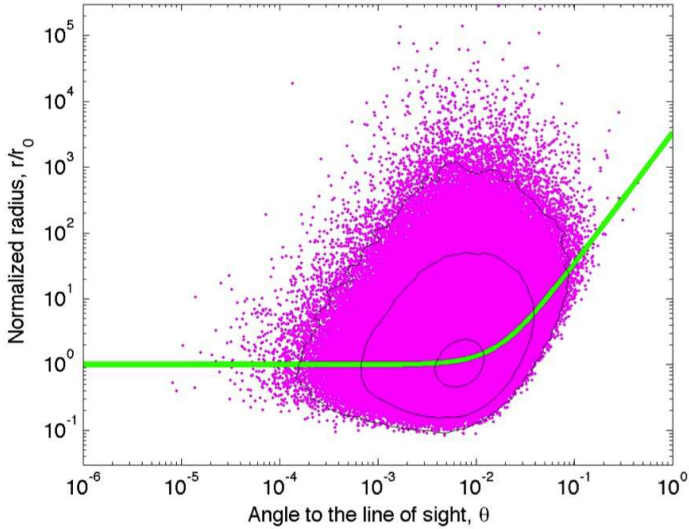}
\caption{The green line represent the (normalized) photospheric radius
  $r_{ph}$ as a function of the angle to the line of sight, $\theta$,
  for spherical explosion (see equation \ref{eq:r_ph_theta}). The red
  dots represent the last scattering locations of photons ejected in
  the center of relativistic expanding ``fireball'' (using a
  Monte-Carlo simulation). The black lines show contours. Clearly,
  photons can undergo their last scattering at a range of radii and
  angles, leading to the concept of ``vague photosphere''. The
  observed photospheric signal is therefore smeared both in time and
  energy.  Figure taken from \cite{Peer08}.  }
\label{fig:photosphere}
\end{figure}

An even closer inspection reveals that photons do not necessarily decouple
the plasma at the photospheric surface; this surface of
$\tau(r,\theta) = 1$ simply represent a probability of $e^{-1}$ for a
photon to decouple the plasma. Instead, the photons have a finite probability of
decoupling the plasma at every location in space. This is demonstrated
in Figure \ref{fig:photosphere}, adopted from [\cite{Peer08}]. This
realization led A. Beloborodov to coin the term ``vague photosphere''
[\cite{Beloborodov11}].

The immediate implication of this non-trivial shape of the photosphere
is that the expected radiative signal emerging from the photosphere
cannot have a pure ``Planck'' shape, but is observed as a gray-body,
due to the different Doppler boosts and different adiabatic energy
losses of photons below $r_{ph}$ [\cite{Peer08, PR11}]. This is in
fact the relativistic extension of the ``limb darkening'' effect known
from stellar physics. As will be discussed in section
\ref{sec:geometry} below, while in spherical outflow only a moderate
modification to a pure ``Planck'' spectra is expected, this effect
becomes extremely pronounced when considering more realistic jet
geometries, and can in fact be used to study GRB jet
geometries [\cite{LPR13}].

\subsection{Relativistic expansion of magnetized outflows} 
\label{sec:magnetized_flow}

\subsubsection{The magnetar model} 

A second type of models assumes that the energy released during the
collapse (or the merger) is not converted directly into
photon-dominated outflow, but instead, is initially used in producing
very strong magnetic fields (Poynting flux dominated plasma). Only at
a second stage, the energy stored in the magnetic field is used in
both accelerating the outflow to relativistic speeds (jet production
and acceleration) as well as heating the particles within the jet.

There are a few motivations for considering this alternative scenario.
Observationally, one of the key discoveries of the Swift satellite is
the existence of a long lasting ``plateau'' seen in the the early
afterglow of GRBs at the X-ray band [\cite{Zhang+06, Nousek+06,
    Rowlinson+13}]. This plateau is difficult to explain in the
context of jet interaction with the environment, but can be explained
by continuous central engine activity (though it may be explained by
other mechanisms, e.g., reverse shock emission; see, [\cite{VE14a,
    VE14b}]) . A second motivation is the fact that magnetic fields
are long thought to play a major role in jet launching in other
astronomical objects, such as AGNs, via the Blandford-Znajek
[\cite{BZ77}] or the Blandford-Payne [\cite{BP82}] mechanisms. These
mechanisms have been recently tested and validated with state of the
art numerical GR-MHD simulations [\cite{Komissarov01, MKU01, MG04,
    Komissarov04, Mckinney05, McKinney06, Komissarov07, TMN08, TNM10,
    TNM11}]; see further explanations in \cite{Spruit10}. It is thus
plausible that they may play some role in the context of GRBs as well.

The key idea is that the core collapse of the massive star does not
form a black hole immediately, but instead leads to a rapidly rotating
proto-neutron star, with a period of $\sim 1$~ms, and very strong
surface magnetic fields ($B \gsim 10^{15}$~G). This is known as the
``magnetar'' model [\cite{Usov92, Thompson94, KR98, Spruit99,
    Wheeler00, TCQ04}]. The maximum energy that can be stored in a
rotating neutron star is $\sim 2 \times 10^{52}$~erg, and the typical
timescale over which this energy can be extracted is $\sim 10$~s (for
this value of the magnetic field). These value are similar to the
values observed in long GRBs. The magnetic energy extracted drives a
jet along the polar axis of the neutron star [\cite{UM07,
  Bucciantini+08, Bucciantini+09, Komissarov+09, GL13, GL14}]. Following this main
energy extraction, residual rotational or magnetic energy may continue
to power late time flaring or afterglow emission, which may be the
origin of the observed X-ray plateau [\cite{Metzger+11}].

\subsubsection{Scaling laws for jet acceleration in magnetized outflows}

Extraction of the magnetic energy leads to acceleration of particles
to relativistic velocities.  The evolution of the hydrodynamic
quantities in these Poynting-flux dominated outflow was considered by
several authors [\cite{SDD01, Drenkhahn02, DS02, VK03, Giannios05,
    Giannios06, GS05, MR11}]. The scaling laws of the acceleration can
be derived by noting that due to the high baryon load, ideal MHD limit
can be assumed [\cite{SDD01}].

In this model, there are two parts to the luminosity
[\cite{Drenkhahn02}]: a kinetic part, $L_k = \Gamma \dot M c^2$, and a
magnetic part, $L_M = 4 \pi r^2 c \beta (B^2/4 \pi)$, where $\beta$ is
the outflow velocity. Thus, $L = L_k + L_M$. Furthermore in this
model, throughout most of the jet evolution the dominated component of
the magnetic field is the toroidal component, and so $\vec B \perp
\vec \beta$.
 
An important physical quantity is the magnetization
parameter, $\sigma$, which is the ratio of Poynting flux to kinetic
energy flux:
\beq
\sigma \equiv {L_M \over L_k} =  {B^2 \over 4 \pi \Gamma^2 n m_p c^2}
\eeq
At the Alfv\'en radius, $r_0$ (at $r=r_0$, the flow velocity is equal
to the Alfv\'en speed), the key assumption is that the flow is highly
magnetized, and so the magnetization is $\sigma(r_0) \equiv \sigma_0
\gg 1$. The magnetization plays a similar role to that of the baryon
loading, in the classical fireball model.

The basic idea is that the magnetic field in the flow changes polarity
on a small scale, $\lambda$, which is of the order of the light
cylinder in the central engine frame ($\lambda \approx 2 \pi
c/\Omega$), where $\Omega$ is the angular frequency of the central
engine - either a spinning neutron star or black hole; see
[\cite{Coroniti90}]. This polarity change leads to magnetic energy
dissipation via reconnection process. It is assumed that the
dissipation of magnetic energy takes place at a constant rate, that is
modeled by a fraction $\epsilon$ of the Alfv\'en speed. As the details
of the reconnection process are uncertain, the value of $\epsilon$ is
highly uncertain. Often a constant value $\epsilon \approx 0.1$ is
assumed. This implies that the (comoving) reconnection time is
$t'_{rec} \sim \lambda'/\epsilon v'_A$, where $v'_A$ is the (comoving)
Alfv\'en speed, and $\lambda' = \Gamma \lambda$. Since the plasma is
relativistic, $v'_A \sim c$, and one finds that $t'_{rec} \propto
\Gamma$. In the lab frame, $t_{rec} = \Gamma t'_{rec} \propto
\Gamma^2$.

Assuming that a constant fraction of the dissipated magnetic energy is
used in accelerating the jet, the rate of kinetic energy increase is
therefore given by
\beq
{d E_k \over dr} \propto {d\Gamma \over dr} \sim {1 \over c t_{rec}}
\propto \Gamma^{-2},
\eeq
from which one immediately finds the scaling law $\Gamma(r) \propto
r^{1/3}$.

The maximum Lorentz factor that can be achieved in this mechanism is
calculated as follows. First, one writes the total luminosity as $L =
L_k + L_M = (\sigma_0 + 1) \Gamma_0 \dot M c^2$, where $\Gamma_0$ is
the Lorentz factor of the flow at the Alfv\'en radius.  Second,
generalization of the Alfv\'enic velocity to relativistic speeds
[\cite{Lich67, Gedalin93}] reads
\beq
\gamma_A \beta_A = {B' \over \sqrt{4 \pi n m_p c^2}} = {B /\Gamma
  \over \sqrt{4 \pi n m_p c^2}} = \sqrt{\sigma}
\eeq
By definition of the Alfv\'enic radius, the flow Lorentz factor at
this radius is $\Gamma_0 = \gamma_A \simeq \sqrt{\sigma_0}$ (since at
this radius the flow is Poynting-flux dominated, $\sigma_0 \gg 1$).
Thus, the mass ejection rate is written as $\dot M \approx
L/\sigma_0^{3/2} c^2$. As the luminosity is assumed constant
throughout the outflow, the maximum Lorentz factor is reached when $L
\sim L_k \gg L_M$, namely $L = \Gamma_{\max} \dot M c^2$. Thus,
\beq
\Gamma_{\max} \approx \sigma_0^{3/2}.
\eeq

In comparison to the photon-dominated outflow, jet acceleration in the
Poynting-flux dominated outflow model is thus much more gradual. The
saturation radius is at $r_s = r_0 \sigma_0^3 \approx
10^{13.5}\,\sigma_2^3 (\epsilon \Omega)_3^{-1}$~cm. Similar
calculations to that presented above show the photospheric radius to
be at radius [\cite{GS05}]
\beq
r_{ph} = 6 \times 10^{11} {L_{52}^{3/5} \over (\epsilon
  \Omega)_3^{2/5} \sigma_2^{3/2}}~{\rm cm}.
\eeq
which is similar (for the values of parameters chosen) to the
photospheric radius obtained in the photon-dominated flow. Note that
in this scenario, the photosphere occurs while the flow is still
accelerating.

The model described above is clearly very simplistic. In particular,
it assumes constant luminosity, and constant rate of reconnection
along the jet. As such, it is difficult to explain the observed rapid
variability in the framework of this model. Furthermore, one still
faces the need to dissipate the kinetic energy in order to produce the
observed $\gamma$-rays. As was shown by several authors [\cite{ZK05,
    MGA09, MA10}], kinetic energy dissipation via shock waves is much
less efficient in Poynting-flow dominated plasma relative to weakly
magnetized plasma.

Moreover, even if this is the correct model in describing (even if
only approximately) the magnetic energy dissipation rate, it is not
known what fraction of the dissipated magnetic energy is used in
accelerating the jet (increasing the bulk Lorentz factor), and what
fraction is used in heating the particles (increasing their random
motion). Lacking clear theoretical model, it is often simply assumed
that about half of the dissipated energy is used in accelerating the
jet, the other half in heating the particles [\cite{SD04}]. Clearly,
all these assumptions can be questioned. Despite numerous efforts in
recent years in studying magnetic reconnection [e.g., \cite{UM11,
    MU12, Cerutti+12, Cerutti+13b, Werner+14}] this is still an open
issue.

Being aware of these limitations, in recent years several authors have
dropped the steady assumption, and considered models in which the
acceleration of a magnetic outflow occurs over a finite, short
duration [\cite{Cont95, TNM10b, Komissarov+10, Granot+11}]. The basic
idea is that variability in the central engine leads to the ejection
of magnetized plasma shells, that expand due to internal magnetic
pressure gradient once they lose causal contact with the source. 

One suggestion is that similar to the internal shock model, the shells
collide at some radius $r_{coll}$. The collision distort the ordered
magnetic field lines entrained in the ejecta. Once reaching a critical
point, fast reconnection seeds occur, which induce relativistic MHD
turbulence in the interaction regions. This model, known as
Internal-Collision-induced Magnetic Reconnection and Turbulence
(ICMART) [\cite{ZY11}] may be able to overcome the low efficiency
difficulty of the classical internal shock scenario.

\subsection{Particle acceleration}
\label{sec:acc}

In order to produce the non-thermal spectra observed, one can in
principle consider two mechanisms. The first is emission of radiation
via various non-thermal processes, such as synchrotron, Compton,
etc. This is the traditional way which is widely considered in the
literature. A second way which was discussed only recently is the use
of light aberration, to modify the (naively expected) Planck spectrum
emitted at the photosphere. The potentials and drawbacks of this
second idea will be considered in section \ref{sec:geometry}. First,
let me consider the traditional way of producing the spectra via
non-thermal radiative processes\footnote{A photospheric emission
  cannot explain photons at the GeV range, and thus even if it does
  play a major role in producing the observed spectra, it is
  certainly not the only radiative mechanism.}.

The internal collisions, magnetic reconnection, or possibly other
unknown mechanism dissipate part of the outflow kinetic
energy\footnote{Within the context of Poynting-flux dominated
  outflows, it was suggested by [\cite{LB03, Lyutikov06}] that the
  magnetic energy dissipated may be converted directly into radiating
  particles, without conversion to kinetic energy first.}. This
dissipated energy, in turn, can be used to heat the particles (increase
their random motion), and/or accelerate some fraction of them to a
non-thermal distribution. Traditionally, it is also assumed that some
fraction of this dissipated energy is used in producing (or enhancing)
magnetic fields. Once accelerated, the high energy particles emit the
non-thermal spectra.

\begin{figure}
\includegraphics[width=10cm]{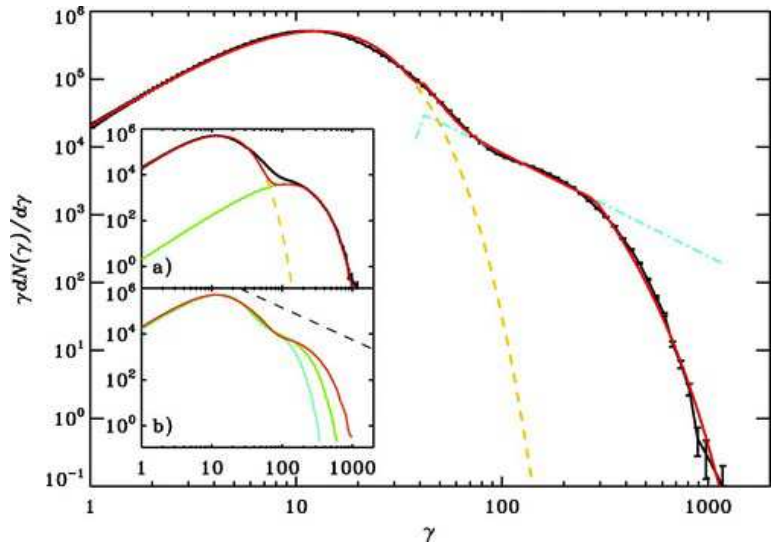}
\caption{Results of a particle in cell (PIC) simulation shows the
  particle distribution downstream from the shock (black line). The
  red line is a fit with a low energy Maxwellian, and a high energy
  power law, with a high energy exponential cutoff. Sub-panel $a$ is
  the fit with a sum of high and low temperature Maxwellians (red
  line), showing a deficit at intermediate energies; subpanel $b$ is
  the time evolution of a particle spectrum in a downstream slice at
  three different times.  The black dashed line shows a
  $\gamma^{-2.4}$ power law. Figure taken from \cite{Spit08}.  }
\label{fig:acc}
\end{figure}

The most widely discussed mechanism for acceleration of particles is
the {\it Fermi mechanism} [\cite{Fermi49, Fermi54}], which requires
particles to cross back and forth a shock wave.  Thus, this mechanism
is naturally associated with internal shell collisions, where shock
waves are expected to form. A basic explanation of this mechanism can
be found in the textbook by [\cite{Longair11}]. For reviews see
[\cite{Bell78, BO78, BE87, JE91}]. In this process, the accelerated
particle crosses the shock multiple times, and in each crossing its
energy increases by a (nearly) constant fraction, $\Delta E / E \sim
1$. This results in a power law distribution of the accelerated
particles, $N(E) \propto E^{-S}$ with power law index $S \approx 2.0 -
2.4$ [\cite{Kirk+98, Kirk+00, Ellison+90, Achterberg+01, ED04}].
Recent developments in particle-in-cell (PIC) simulations have allowed
to model this process from first principles, and study it in more
detail [\cite{Silva+03, Nishikawa+03, Spit08, SS09b, Haug11}]. As can
be seen in Figure \ref{fig:acc} taken from [\cite{Spit08}], indeed a
power law tail above a low energy Maxwellian in the particle
distribution is formed.

The main drawback of the PIC simulations is that due to the numerical
complexity of the problem, these simulations can only cover a tiny
fraction ($\sim 10^{-8}$) of the actual emitting region in which
energetic particles exist. Thus, these simulations can only serve as
guidelines, and the problem is still far from being fully
resolved. Regardless of the exact details, it is clear that particle
acceleration via the Fermi mechanism requires the existence of shock
waves, and is thus directly related to the internal dynamics of the
gas, and possibly to the generation of magnetic fields, as mentioned
above.

\begin{figure}
\includegraphics[width=7cm]{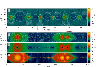}
\includegraphics[width=7cm]{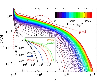}
\caption{Results of an electromagnetic particle in cell (PIC)
  simulation TRISTAN-MP show the structure of the reconnection layer
  (left) and the accelerated particle distribution function
  (right). Left: structure of the reconnection layer. Shown are the
  particle densities (a), (b); magnetic energy fraction (c) and mean
  kinetic energy per particle (d). The plasmoids are clearly
  seen. Right: temporal evolution of particle energy spectrum. The
  spectrum at late times resembles a power law with slope $p=2$
  (dotted red line), and is clearly departed from a Maxwellian. The
  dependence of the spectrum on the magnetization is shown in the
  inset. Figure is takes from \cite{SS14}.}
\label{fig:acc2}
\end{figure}

The question of particle acceleration in magnetic reconnection layers
have also been extensively addressed in recent years [see \cite{RL92,
    ZE01, Lyutikov03, Jaroschek+04, Lyubarsky05, ZE07, ZE08, LL08,
    Yin+08, Giannios10b, Lazarian+11, Liu+11, UM11, MU12, BB12,
    Cerutti+12, Cerutti+13b, Kagan+13, Werner+14, SS14, US14} for a
  partial list of works].  The physics of acceleration is somewhat
more complicated than in non-magnetized outflows, and may involve
several different mechanisms. The basic picture is that the
dissipation of the magnetic field occurs in sheets. The first mechanism
relies on the realization that within these sheets, there are regions
of high electric fields; particles can therefore be accelerated
directly by the strong electric fields. A second mechanism is based on
instabilities within the sheets, that create ``magnetic islands'',
(plasmoids) that are moving close to the Alfv\'en speed (see Figure
\ref{fig:acc2}). Particles can therefore be accelerated via Fermi
mechanism by scattering between the plasmoids. A third mechanism is
based on converging plasma flows towards the current sheets, that
provide another way of particle acceleration via first order Fermi
process.

In addition, if the flow is Poynting-flux dominated, particles may
also be accelerated in shock waves; however, it was argued that
Fermi-type acceleration in shock waves that may develop in highly
magnetized plasma may be inefficient [\cite{SS09b, SS11}]. Thus, while
clearly addressing the question of particle acceleration in magnetized
outflow is a very active research field, the numerical limitations
imply that theoretical understanding of this process, and its details
(e.g., what fraction of the reconnected energy is being used in
accelerating particles, or the energy distribution of the accelerated
particles) is still very limited.

Although the power law distribution of particles resulting from
Fermi-type, or perhaps magnetic-reconnection acceleration is the most
widely discussed, we point out that alternative models exist. One such
model involves particle acceleration by a strong electromagnetic
potential, which can exceed $10^{20}$~eV close to the jet core
[\cite{Lovelace76, Blandford76, Neronov+09}]. The accelerated
particles may produce a high energy cascade of electron-positron
pairs. Additional model involves stochastic acceleration of particles
due to resonant interactions with plasma waves in the black hole
magnetosphere [\cite{Dermer+96}].

Several authors have also considered the possibility that particles in
fact have a relativistic quasi-Maxwellian distribution [\cite{JH79,
    CJ80, WZ00, PC09}]. Such a distribution, with the required
temperature ($\sim 10^{11} - 10^{12}$~K) may be generated if particles
are roughly thermalized behind a relativistic strong shock wave [e.g.,
  \cite{BM77}]. While such a model is consistent with several key
observations, it is difficult to explain the very high energy (GeV)
emission without invoking very energetic particles, and therefore some
type of particle acceleration mechanism must take place as part of the
kinetic energy dissipation process.

\subsection{Radiative processes and the production of the observed spectra.}
\label{sec:radiation}

Following jet acceleration, kinetic energy dissipation (either via
shock waves or via magnetic reconnection) and particle acceleration,
the final stage of energy conversion must produce the observed
spectra. As the $\gamma$-ray spectra is both very broad and
non-thermal (does not resemble ``Planck''), most efforts to date are
focused on identifying the relevant radiative processes and physical
conditions that enable the production of the observed spectra. The
leading radiative models initially discussed are synchrotron emission,
accompanied by synchrotron-self Compton at high energies. However, as
has already mentioned, it was shown that this model is inconsistent
with the data, in particular low energy spectral slopes.

Various suggestions of ways to overcome this drawback by modifying
some of the physical conditions and / or physical properties of the
plasma were proposed in the last decade. However, a major revolution
occurred with the realization that part of the spectra is thermal. This
led to new set of models in which part of the emission originates from
below the photosphere (the optically thick region). It should be
stressed that only part of the spectrum - but not all of it is assumed
to originate from the photosphere. Thus, in these models as well,
there is room for optically thin (synchrotron and IC) emission,
originating from a different location. Finally, a few most recent works
on light aberration show that the contribution of the photospheric
emission may be much broader than previously thought.

\subsubsection{Optically thin model: synchrotron}

Synchrotron emission is perhaps the most widely discussed model for
explaining GRB prompt emission. It has several advantages. First, it
has been extensively studied since the 1960's [\cite{GS65, BG70}] and
is the leading model for interpreting non-thermal emission in AGNs,
XRBs and emission during the afterglow phase of GRBs.  Second, it is
very simple: it requires only two basic ingredients, namely energetic
particles and a strong magnetic field. Both are believed to be
produced in shock waves (or magnetic reconnection phase), which tie it
nicely to the general ``fireball'' (both ``hot'' and ``cold'') picture
discussed above. Third, it is broad band in nature (as opposed, e.g.,
to the ``Planck'' spectrum), with a distinctive spectral peak, that
could be associated with the observed peak energy. Fourth, it provides
a very efficient way of energy transfer, as for the typical
parameters, energetic electrons radiate nearly 100\% of their energy.
These properties made synchrotron emission the most widely
discussed radiative model in the context of GRB prompt emission
[\cite{RM92, MR93, MLR93, MRP94, PX94, PM96, Tavani96, Cohen+97,
    SP97b, PL98, DM98} for a very partial list].

Consider a source at redshift $z$ which is moving
at velocity $\beta \equiv v/c$ (corresponding Lorentz factor
$\Gamma = (1-\beta)^{-1/2}$) at angle $\theta$ with respect to the
observer. The emitted photons are thus seen with a Doppler boost $\D =
[\Gamma(1-\beta \cos\theta)]^{-1}$. Synchrotron emission from electrons
having random Lorentz factor $\gamma_{el}$ in a magnetic field $B$ (all
in the comoving frame) is observed at a typical energy
\beq 
\varepsilon_m^{ob} = {3 \over 2} \hbar {q B \over m_e c}
\gamma_{el}^2 {\D \over (1 +z)} = 1.75 \times 10^{-19} B \gamma_{el}^2
      {\D \over (1+ z)} \erg.
\label{eq:nu_m}
\eeq

If this model is to explain the peak observed energy, $\epsilon^{ob}
\approx 200$~keV with typical Lorentz factor $\D \simeq \Gamma \sim
100$ (relevant for on-axis observer), one obtains a condition on the
typical electron Lorentz factor and magnetic field,
\beq
B \gamma_{el}^2 \simeq 3.6 \times 10^{10} \left({1 + z \over 2}\right)
\Gamma_2^{-1}\, \left({\epsilon^{ob} \over 200~\keV}\right) ~{\rm{G}}
\label{eq:Bg}
\eeq
Thus, both strong magnetic field and very energetic electrons are
required in interpreting the observed spectral peak as due to
synchrotron emission. Such high values of the electrons Lorentz factor
are not excluded by any of the known models for particle
acceleration. High values of the magnetic fields may be present if the
outflow is Poynting flux dominated. In the photon-dominated outflow,
strong magnetic fields may be generated via two stream (Weibel)
instabilities [\cite{Weibel59, ML99, Silva+03, Frederiksen+04,
    Nishikawa+05, Spit08a}]. 

One can therefore conclude that the synchrotron model is capable of
explaining the peak energy. However, one alarming problem is that the
high values of both $B$ and $\gamma_{el}$ required, when expressed as
fraction of available thermal energy (the parameters $\epsilon_e$ and
$\epsilon_B$) are much higher than the (normalized) values inferred
from GRB afterglow measurements [\cite{WRM97, PK02, SBK14,
    B-D14}]. This is of a concern, since broad band GRB afterglow
observations are typically well fitted with the synchrotron model, and
the microphysics of particle acceleration and magnetic field
generation should be similar in both prompt and afterglow
environments\footnote{Though the forward shock producing the afterglow
  is initially highly relativistic, while shock waves produced during
  the internal collisions may be mildly relativistic at most.}.

The main concern though is the low energy spectral slope. As long as
the electrons maintain their energy, the expected synchrotron spectrum
below the peak energy is $F_\nu \propto \nu^{1/3}$ (corresponding
photon number $N_E \propto E^{-2/3}$) [e.g., \cite{RL79}]. This is
roughly consistent with the observed low energy spectral slope,
$\langle \alpha \rangle = -1$ (see Section \ref{sec:Band}).

However, at these high energies, and with such strong magnetic field,
the radiating electrons rapidly cool by radiating their energy on a
very short time scale:
\beq 
t'_{\rm cool} = {E \over P} = {\gamma_{el} m_e c^2 \over (4/3) c
  \sigma_T \gamma_{el}^2 u_B (1 + Y)} = {6 \pi m_e c \over \sigma_T
  B^2 \gamma_{el} (1 +Y)},
\label{eq:t_cool}
\eeq
Here, $E= \gamma_{el} m_e c^2$ is the electron's energy, $P$ is the
radiated power, $u_B \equiv B^2/8\pi$ is the energy density in the
magnetic field, $\sigma_T$ is Thomson's cross section and $Y$ is
Compton parameter. The factor ($1+Y$) is added to consider cooling via
both synchrotron and Compton scattering. 

Using the values obtained in Equation \ref{eq:Bg}, one finds the
(comoving) cooling time to be 
\beq
t'_{\rm cool} = 6.0 \times 10^{-13} \gamma_{el}^3 ~ (1+Y)^{-1}
\left({1 + z \over 2}\right)^{-2} \Gamma_2^2\, \left({\epsilon^{ob}
  \over 200~\keV}\right)^{-2}~{\rm s}.
\eeq
This time is to be compared with the comoving dynamical time,
$t'_{dyn} \sim R/\Gamma c$. If the cooling time is shorter than the
dynamical time, the resulting spectra below the peak is $F_\nu \propto
\nu^{-1/2}$ [e.g., \cite{SNP96, SPN98}], corresponding to $N_E
\propto E^{-3/2}$. While values of the power law index {\it smaller} than
$-3/2$, corresponding to shallow spectra can be obtained by
superposition of various emission sites, steeper values cannot be
obtained. Thus, the observed low energy spectral slope of $\sim85\%$
of the GRBs (see Figure \ref{fig:BAND_distribution}) which show
$\alpha$ larger than this value ($\langle \alpha \rangle = -1$) cannot
be explained by synchrotron emission model. This is the ``synchrotron
line of death'' problem introduced above.

The condition for $t'_{\rm cool} \gsim t'_{dyn}$ can be written as 
\beq
\gamma_{el} \gsim 3.8 \times 10^{4} R_{14}^{1/3} (1+Y)^{1/3} \left({1
  + z \over 2}\right)^{2/3}\, \Gamma_2^{-1} \, \left({\epsilon^{ob}
  \over 200~\keV}\right)^{2/3}.  
\eeq
The value of the emission radius $R=10^{14}$~cm is chosen as a
representative value that enables variability over time scale $\delta
t^{ob} \sim R/\Gamma^2 c \sim 0.3~ R_{14} \Gamma_2^{-2}$~s.

Since $\gamma_{el}$ represents the characteristic energy of the
radiating electrons, such high values of the {\it typical} Lorentz
factor $\gamma_{el}$ are very challenging for theoretical
modeling. However, a much more severe problem is that in this model,
under these conditions, the energy content in the magnetic field must
be very low (see Equation \ref{eq:Bg}). In order to explain the
observed flux, one must therefore demand high energy content in the
electron's component, which is several orders of magnitude higher than
that stored in the magnetic field [\cite{KM08, BP13, KZ14}]. This, in
turn, implies that inverse Compton becomes significant, producing
$\sim$~TeV emission component that substantially increase the total
energy budget. As was shown by [\cite{KM08}], such a scenario can only
be avoided if the emission radius is $R \gsim 10^{17}$~cm, in which
case it is impossible to explain the rapid variability observed. Thus,
the overall conclusion is that classical synchrotron emission as a
leading radiative process fails to explain the key properties of the
prompt emission of the vast majority of GRBs [\cite{Ghisellini+00b,
    Preece+02}].

\subsubsection{Suggested modifications to the classical synchrotron scenario}

The basic synchrotron emission scenario thus fails to
self-consistently explain both the energy of the spectral peak and the
low energy spectral slope. In the past decade there have been several
suggestions of ways in which the basic picture might be modified, so
that the modified synchrotron emission, accompanied by inverse Compton
scattering of the synchrotron photons (synchrotron-self Compton; SSC)
would be able to account for these key observations.

The key problem is the fast cooling of the electrons, namely $t_{cool}
< t_{dyn}$. However, in order for the electrons to rapidly cool they
must be embedded in a strong magnetic field. The spatial structure of
the magnetic field is not clear at all. Thus, it was proposed by
[\cite{PZ06}] that the magnetic field may decay on a relatively short
length scale, and so the electrons would not be able to efficiently
cool. This idea had gain interest recently [ \cite{Zhao+14,
    UZ14}]. Its major drawback is the need for high energy budget, as
only a small part of the energy stored in the electrons is radiated.

Another idea is that synchrotron self absorption may produce steep low
energy slope below the observed peak [\cite{LP00}]. However, this
requires unrealistically high magnetic field. Typically, the
synchrotron self absorption frequency is expected at the IR/Optic band
[e.g., \cite{RL79, Granot+00}]. Thus, synchrotron self absorption may
be relevant in shaping the spectrum at the X-rays only under very extreme
conditions [e.g., \cite{PW04b}].

Looking into a different parameter space region, it was suggested that
the observed peak energy is not due to synchrotron emission, but due
to inverse-Compton scattering of the synchrotron photons, which are
emitted at much lower energies [\cite{PM00, DB00, SP04}]. In these
models, the steep low energy spectral slope can result from
up-scattering of synchrotron self absorbed photons. However, a careful
analysis of this scenario (e.g., [\cite{KZ14}]) reveals requirements
on the emission radius, $R \gsim 10^{16}$~cm and optical flux
(associated with the synchrotron seed photons) that are inconsistent
with observations. Furthermore, a second scattering would lead to
substantial TeV flux, resulting in an energy crisis
[\cite{Derishev+01, Piran+09}]. Thus, this model as well is concluded
as not being viable as the leading radiative model during the GRB
prompt emission [\cite{Piran+09}].

If the energy density in the photon field is much greater than in the
magnetic field, then electron cooling by inverse Compton scattering
the low energy photons dominated over cooling by synchrotron
radiation. The most energetic electrons cool less efficiently due to
the Klein-Nishina (KN) decrease in the scattering cross section. Thus,
in this parameter space where KN effect is important, steeper low
energy spectral slopes can be obtained [\cite{Derishev+01, Nakar+09,
    Daigne+11}]. However, even under the most extreme conditions, the
steepest slope that can be obtained is no harder than $F_\nu \sim
\nu^0$ [\cite{Nakar+09, BBK12}], corresponding to $N_E \propto
E^{-1}$ - which can explain at most $\sim 50\%$ of the low energy
spectral slopes observed. Moreover, very high values of the Lorentz
factor, $\gamma_{el} \gsim 10^6$ are assumed which challenge
theoretical models, as discussed above.

A different proposition was that the heating of the electrons may be
slow; namely, the electrons may be continuously heated while radiating
their energy as synchrotron photons. This way, the rapid electrons
cooling is avoided, and a shallower spectra can be obtained
[\cite{GC99, GC99b, KM08, AT09, Murase+12b}]. While there is no known
mechanism that could continuously heat the electrons as they cross the
shock wave and are advected downstream in the classical internal
collision scenario, it was proposed that slow heating may result from
MHD turbulence down stream of the shock front
[\cite{Murase+12b}]. Thus this may be an interesting alternative,
though currently there are still large gaps in the physics involved in
the slow heating process.

A different suggestion is emission by the hadrons (protons). The key
idea is that whatever mechanism that is capable of accelerating
electrons to high energies, should accelerate protons as well; in
fact, the fact that high energy cosmic rays are observed necessitate
the existence of such a mechanism, although its detailed in the
context of GRBs are unknown. Many authors have considered possible
contribution of energetic protons to the observed spectra [e.g.,
  \cite{BD98, Totani98, GZ07, Asano+09, Razzaque+10, AM12, CK13}].
Energetic proton contribution to the spectrum is both via direct
synchrotron emission, and also indirectly by photo-pion production or
photo-pair production.

Clearly, proton acceleration to high energies would imply that GRBs
are potentially strong source of both high energy cosmic rays and
energetic neutrinos [\cite{MU95, W95, WB97, W04}]. On the other hand,
the main drawback of this suggestion is that protons are much less
efficient radiators than electrons (as the ratio of proton to electron
cross section for synchrotron emission $\sim (m_e/m_p)^2$). Thus, in
order to produce the observed luminosity in $\gamma$-rays, the energy
content of the protons must be very high, with proton luminosity of
$\sim 10^{55}- 10^{56}$~erg~s$^{-1}$. This is at least 3 orders of
magnitude higher than the requirement for leptonic models.

\subsubsection{Photospheric emission}
\label{sec:phot_emission}

 As discussed above, photospheric (thermal) emission is an inherent
 part of both the ``hot'' and ``cold'' (magnetized) versions of the
 fireball model. Thus, it is not surprising that the very early models
 of cosmological GRBs considered photospheric emission as a leading
 radiative mechanism [\cite{Goodman86, Pac86,Pac90,
     Thompson94}]. However, following the observational evidence of a
 non-thermal emission, and lacking clear evidence for a thermal
 component, this idea was abandoned for over a decade.

Renewed interest in this idea began in the early 2000's, with the
realization that the synchrotron model - even after being modified,
cannot explain the observed spectra. Thus, several authors considered
addition of thermal photons to the overall non-thermal spectra, being
either dominant [\cite{EL00, DM02}] or sub-dominant [\cite{MR00,
    MRRZ02, RM05}]. Note that as neither the internal collision or the
magnetic reconnection models provide clear indication of the location
and the amount of dissipated kinetic energy that is later converted
into non-thermal radiation, it is impossible to determine the expected
ratio of thermal to non-thermal photons from first principles in the
framework of these models. Lacking clear observational evidence, it
was therefore thought that $r_{ph} \gg r_s$, in which case adiabatic
losses lead to strong suppression of the thermal luminosity and
temperature (see Equations \ref{eq:T_ob}, \ref{eq:L_th}).

However, as was pointed out by [\cite{PW04}], in the scenario where
$r_{ph} \gg r_s$ it is possible that substantial fraction of kinetic
energy dissipation occurs below the photosphere (e.g., in the internal
collision scenario, if $r_{coll} < r_{ph}$). In this case, the
radiated (non-thermal) photons that are emitted as a result of the
dissipation process cannot directly escape, but are advected with the
flow until they escape at the photosphere. This triggers several
events. First, multiple Compton scattering substantially modifies the
optically thin (synchrotron) spectra, presumably emitted initially by
the heated electrons. Second, the electrons in the plasma rapidly
cool, mainly by IC scattering. However, they quickly reach a 'quasi
steady state', and their distribution becomes quasi-Maxwellian,
irrespective of their initial (accelerated) distribution. The
temperature of the electrons is determined by balance between heating
- both external, as well as by direct Compton scattering energetic
photons, and cooling (adiabatic and radiative) [\cite{PMR05}]. The
photon field is then modified by scattering from this quasi-Maxwellian
distribution of electrons. The overall result is a regulation of the
spectral peak at $\sim 1$~MeV (for dissipation that takes place at
moderate optical depth, $\tau \sim$a few - few tens), and low energy
spectral slopes consistent with observations [\cite{PW04}].

\begin{figure}
\includegraphics[width=10cm]{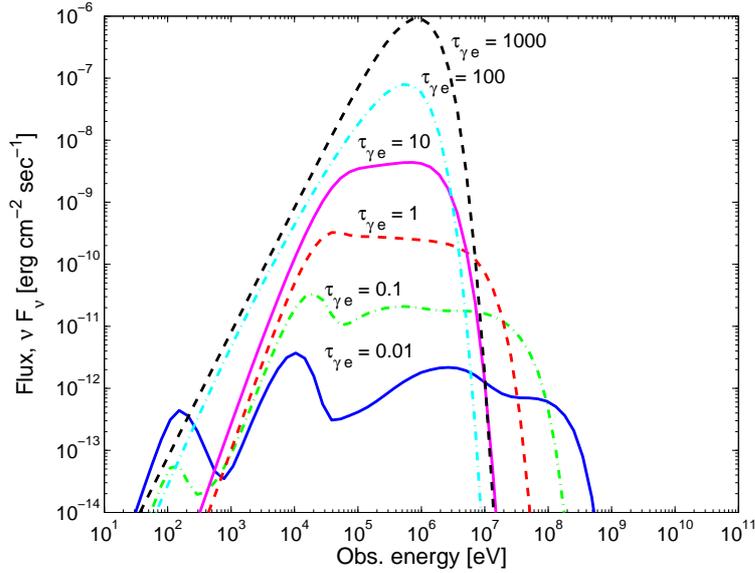}
\caption{Time averaged broad band spectra expected following kinetic
  energy dissipation at various optical depths. For low optical depth,
  the two low energy bumps are due to synchrotron emission and the
  original thermal component, and the high energy bumps are due to
  inverse Compton phenomenon.  At high optical depth, $\tau
  \geq 100$, a Wien peak is formed at $\sim 10 \keV$, and is
  blue-shifted to the MeV range by the bulk Lorentz factor $\simeq
  100$ expected in GRBs.  In the intermediate regime, 
$0.1 < \tau < 100$, 
a flat energy spectrum above the thermal
  peak is obtained by multiple Compton scattering. Figure is takes
  from \cite{PMR06}.}
\label{fig:PMR06}
\end{figure}

The addition of the thermal photons that originate from the initial
explosion (this term is more pronounced if $r_{ph} \gsim r_s$)
significantly enhances these effects [\cite{PMR06}]. The thermal
photons serve as seed photons for IC scattering, resulting in rapid
cooling of the non-thermal electrons that are heated in the
sub-photospheric energy dissipation event. As the rapid IC cooling
leads to a quasi-steady state distribution of the electrons, the
outcome is a 'two temperature plasma', with electron temperature
higher than the thermal photon temperature, $T_{el} > T_{ph}$. An
important result of this model is that the electron temperature is
highly regulated, and is very weakly sensitive to the model
uncertainties; see [\cite{PMR05}] for details. If the dissipation
occurs at intermediate optical depth, $\tau \sim$few - few tens, the
emerging spectrum has a nearly 'top hat' shape (see Figure
\ref{fig:PMR06}). Below $T_{ph}$ the spectrum is steep, similar to the
Rayleigh-Jeans part of the thermal spectrum; in between $T_{ph}$ and
$T_{el}$, a nearly flat energy spectra, $\nu f_\nu \propto \nu^0$
(corresponding $N_E \propto E^{-2}$) is obtained, resulting from
multiple Compton scattering; and an exponential cutoff is expected at
higher energies.

Interestingly, the spectral slope obtained in the intermediate regime
is similar to the obtained high energy spectral slope, $\langle \beta
\rangle \sim -2$ (see discussion in section \ref{sec:Band} and Figure
\ref{fig:BAND_distribution}). Thus, a simple interpretation is to
associate the observed $E_{pk}$ with $T_{ph}$. However, this is likely
a too simplistic interpretation from the following reasons. First, the
predicted low energy spectral slopes, being (modified) thermal are
typically {\it harder} than the observed [\cite{DZ14}]. Second, in
GRB110721A, the peak energy is at $\approx 15$~MeV at early times
[\cite{Axelsson+12, Iyyani+13}], which is too high to be accounted for
by $T_{ph}$ [\cite{Zhang+12, Veres+12}]. Moreover, recent analysis of
Fermi data show a thermal peak at lower energies than $E_{pk}$ (see,
e.g., figure \ref{fig:Iyyani}), which is consistent with the
interpretation of the thermal peak being associated with $T_{ph}$. The
key result of this model, that $T_{ph} < T_{el}$ is consistent with
the observational result of $E_{peak, th} < E_{pk}$, which is
applicable to all GRBs in which thermal emission was identified so far.
This model thus suggests that $E_{pk}$ may be associated with
$T_{el}$, though it does not exclude synchrotron origin for $E_{pk}$;
see further discussion below.

If the optical depth in which the kinetic energy dissipation takes
place is $\tau \gsim 100$, the resulting spectra is close to thermal;
while if $\tau \lsim$~a few, the result is a complex spectra, with
synchrotron peak, thermal peak and at least two peaks resulting from
IC scattering (see Figure \ref{fig:PMR06}). Below the thermal peak,
the main contribution is from synchrotron photons, that are emitted by
the electrons at the quasi steady distribution. Above the thermal
peak, multiple IC scattering is the main emission process, resulting
in nearly flat energy spectra. Thus, this model naturally predicts
different spectral slopes below and above the thermal peak. 

 Interestingly, the key results of this model do not change if one
 considers highly magnetized plasma [\cite{Giannios06, Giannios08,
     Giannios12, VM12, BP14, GZ15}]. Indeed, as this model of
 sub-photospheric energy dissipation is capable of capturing the key
 observed features of the prompt emission, it attracted a lot of
 attention in recent years [e.g., \cite{Ioka+07, TMR07, Lazzati+09,
     LB10, Beloborodov10, Mizuta+11, LMB11, Toma+11, Bromberg+11,
     Levinson12, Veres+12, VLP13, Beloborodov13, Hascoet+13,
     Lazzati+13, AM13, DZ14, C-M+15, CL15}].

It should be noted that the above analysis holds for a single
dissipation episode. In explaining the complex GRB lightcurve, multiple
such episodes (e.g., internal collisions) are expected. Thus, a
variety of observed spectra, which are superposition of the different
spectra that are obtained by dissipation at different optical depth
are expected [\cite{KL14}].

In spite if this success, this model still suffers two main
drawbacks. The first one already discussed is the need to explain low
energy spectral slopes that are not as hard as the Rayleigh-Jeans part
of a Planck spectra. Further, this model needs to explain the high
peak energy $(>$~MeV) observed in some bursts in a self-consistent
way.  A second drawback is the inability of the sub photospheric
dissipation model to explain the very high energy (GeV) emission
seen. Such high energy photons must originate from some dissipation
above the photosphere.

There are two solutions to these problems. The first is geometric in
nature, and takes into account the non-spherical nature of GRB jets to
explain how low energy spectral slopes are modified. This will be
discussed below. The second is the realization that the photospheric
emission must be accompanied by at least another one dissipation
process that takes place above the photosphere. This conclusion,
however, is aligned with both observations of different temporal
behavior of the high energy component (see section
\ref{sec:high_energy_component}), as well as with the basic idea of
multiple dissipation episodes, inherent to both the ``internal
collision'' model and to the magnetic reconnection model.

\begin{figure}
\includegraphics[width=10cm]{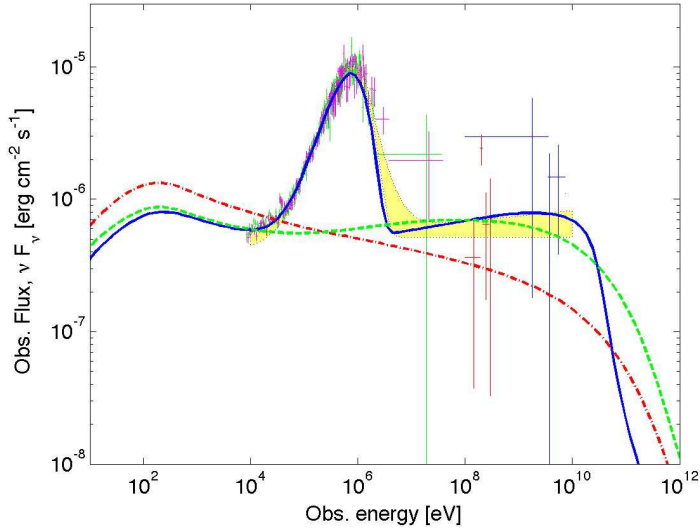}
\caption{Spectral decomposition of GRB090902B (taken at time interval
  (c), 9.6 - 13.0 seconds after the GBM trigger) enables clear
  identification of the physical origin of the emission.  The
  dash-dotted (red) curve shows the spectrum that would have obtained
  if synchrotron radiation was the only source of emission. The dashed
  (green) curve shows the resulting spectrum from synchrotron and SSC,
  and the solid (blue) curve shows the spectrum with the full
  radiative ingredients (synchrotron, SSC, the $\sim$~MeV thermal
  peak, and Comptonization of the thermal photons). Numerical fits are
  done using the radiative code developed by [\cite{PW05}]. Figure is
  takes from \cite{Peer+12}.}
\label{fig:Peer+12}
\end{figure}

Indeed, in the one case in which detailed modeling was done by
considering two emission zones (photosphere and external one), very
good fits to the data of GRB090902B were obtained
[\cite{Peer+12}]. This fits were done with a fully physically
motivated model, which enables to determine the physical conditions at
both emission zones [\cite{Peer+12}]. This is demonstrated in Figure
\ref{fig:Peer+12}. 

\subsubsection{Geometrical Broadening}
\label{sec:geometry}

As was already discussed in section \ref{sec:photosphere}, the
definition of the photosphere as the last scattering surface must be
modified to incorporate the fact that photons have finite probability
of being scattered at every location in space where particles
exist. This led to the concept of ``vague photosphere'' (see Figure
\ref{fig:photosphere}). The observational consequences of this effect
were studied by several authors [\cite{Peer08, Beloborodov10,
    Beloborodov11, LPR13, Ruffini+13, Aksenov+13, Vereschagin14}] In
spherical explosion case, the effect of the vague photosphere is not
large; it somewhat modifies the Rayleigh-Jeans part of the spectrum,
to read $F_{\nu} \propto \nu^{3/2}$ ([\cite{Beloborodov11}]). However,
for non-spherical explosion, the effect becomes dramatic.

While the exact geometry of GRB jets, namely $\Gamma(r,\theta,\phi)$
are unknown, numerical simulations of jets propagating through the
stellar core [e.g., \cite{ZWM03}] suggest a jet profile of the form
$\Gamma(\theta) \sim \Gamma_0/(1+(\theta/\theta_j)^{2p})$, at least
for non-magnetized outflows. Such a jet profile thus assumes a
constant Lorentz factor, $\Gamma \sim \Gamma_0$ for $\theta \lsim
\theta_j$ (the ``jet core'', or inner jet), and decaying Lorentz
factor at larger angles, $\Gamma(\theta) \propto \theta^{-p}$ (outer
jet, or jet sheath). As the Lorentz factor is $\Gamma \propto L/\dot
M$ (\ref{sec:scalings2}), such a profile can result from excess of
mass load close to the jet edge, by mass collected from the star ($\dot
M = \dot M(\theta)$, or alternatively by angle dependent luminosity.

\begin{figure}
\includegraphics[width=8cm]{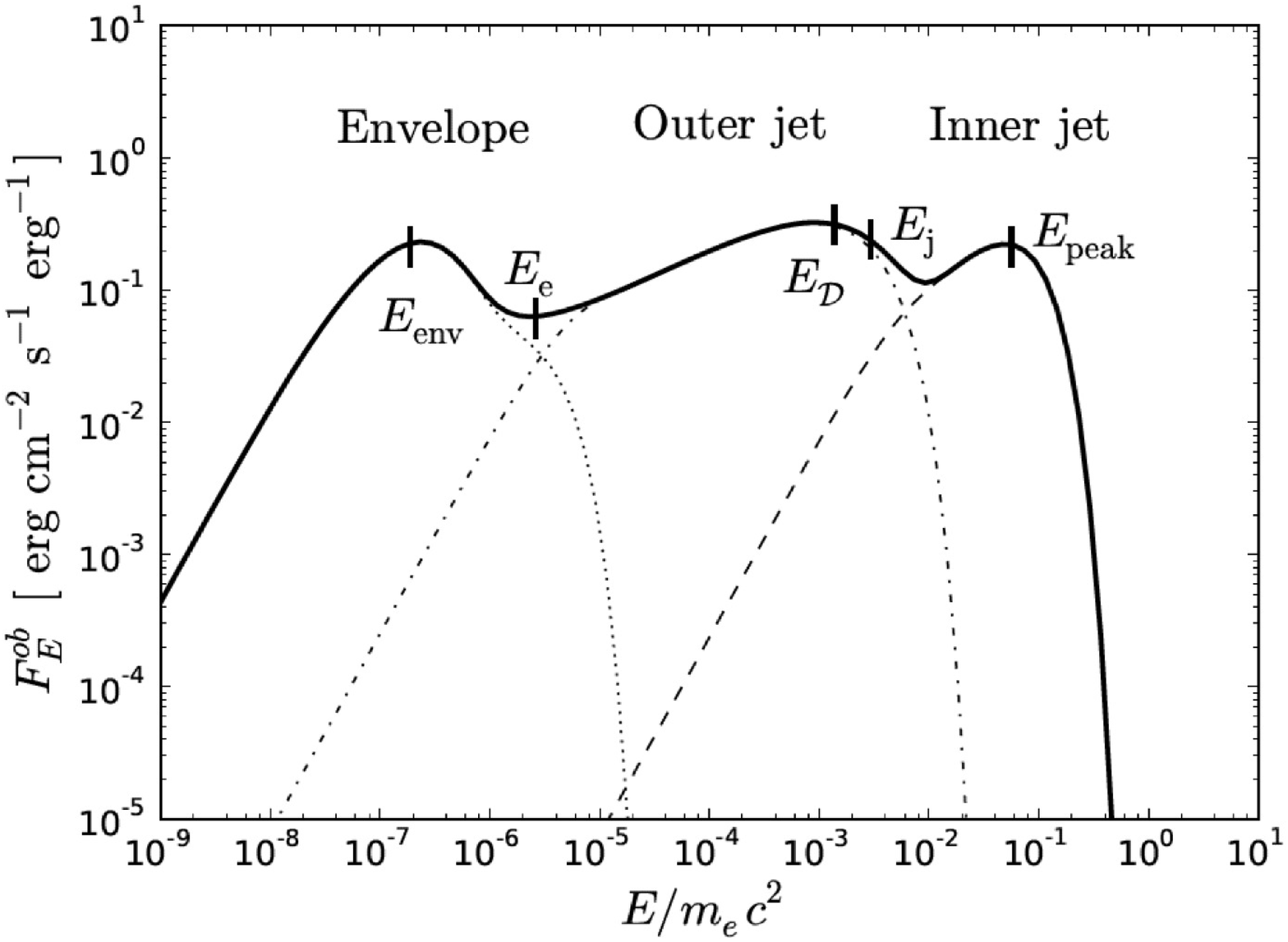}
\includegraphics[width=4cm]{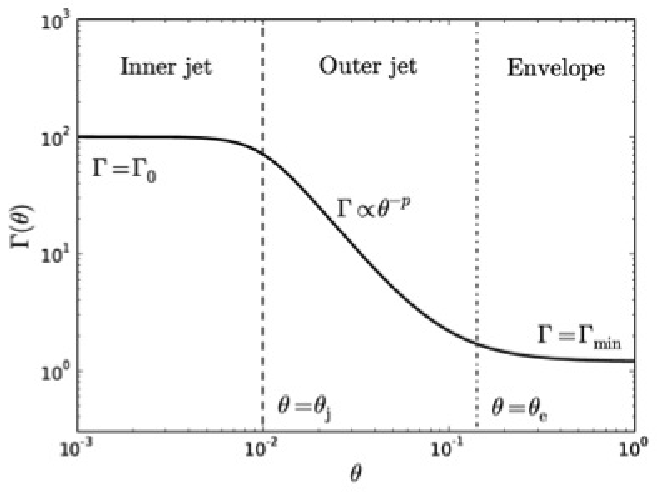}
\caption{{\bf Left.} The expected (observed) spectrum from a
  relativistic, optically thick outflow. The resulting spectra does
  not resemble the naively expected ``Planck'' spectrum. Separate
  integration of the contributions from the inner jet (where $\Gamma
  \approx \Gamma_0$), outer jet (where $\Gamma$ drops with angle) and
  envelope is shown with dashed, dot–dashed and dotted lines,
  respectively. {\bf Right.} The assumed jet profile. Figure taken
  from \cite{LPR13}.  }
\label{fig:Christoffer1}
\end{figure}

The scenario of $\dot M = \dot M(\theta)$ was considered by
[\cite{LPR13}]. While photospheric emission from the inner parts of
the jet result in mild modification to the black body spectrum,
photons emitted from the outer jet's photosphere dominate the spectra
at low energies (see Figure \ref{fig:Christoffer1}). For narrow jets
($\theta_j \Gamma_0 \lsim$~few), this leads to flat low energy spectra,
$dN/dE \propto E^{-1}$, which is independent on the viewing angle, and
very weakly dependent on the exact jet profile. This result thus both
suggests the possibility that the low energy slopes are in fact part
of the photospheric emission, and in addition can be used to infer the
jet geometry.

A second aspect of the model, is that the photospheric emission can be
observed to be highly polarized, with up to $\approx 40\%$
polarization [\cite{LPR14, Chang+14}]. While IC scattering produces
highly polarized light, in spherical models the polarization from
different viewing angles cancels. However, this cancellation is
incomplete in jet-like models (observed off-axis). While the observed
flux by an observer off the jet axis (that can see highly polarized
light) is reduced, it is still high enough to be detected
[\cite{LPR14}].

A third unique aspect that results from jet geometry (rather than
spherical explosion) is photon energy gain by Fermi-like process. As
photons are scattered back and forth between the jet core and the
sheath, on the average they gain energy. This leads to a high energy
power law tail (above the thermal peak) [\cite{LPR13, Ito+13}]. This
again may serve as a new tool in studying jet geometry; though the
importance of this effect in determining the high energy spectra of
GRBs is still not fully clear [Lundman et. al., in prep.].

\subsubsection{A few implications of the photospheric term}

A great advantage of the photospheric emission in its relative
simplicity. By definition, the photosphere is the inner most region
from which electromagnetic signal can reach the observer. Thus, the
properties of the emission site are much more constrained, relative,
e.g., to synchrotron emission (whose emission radius, magnetic field
strength and particle distribution are not known).

In fact, in the framework of the ``hot'' fireball model, the (1-d)
photospheric radius is a function of only two parameters: the
luminosity (which can be measured once the distance is known) and the
Lorentz factor (see Equation \ref{eq:r_ph}). The photospheric radius
is related to the observed temperature and flux via $r_{ph}/\Gamma
\propto (F_{bb}^{ob}/\sigma {T^{ob}}^4)^{1/4}$, where $\sigma$ is
Stefan's constant, and the extra factor of $\Gamma^{-1}$ is due to
light aberration.  Since $r_{ph} \propto L \Gamma^{-3}$, measurements
of the temperature and flux for bursts with known redshift enables an
independent measurement of $\Gamma$, the Lorentz factor at the
photosphere [\cite{Peer+07}]. This, in turn, can be used to determine
the full dynamical properties of the outflow.

One interesting result is that by using this method , it is found that
$r_0$, the size of the jet base, is $\sim 10^{8.5}$~cm, two-three
orders of magnitude above the Schwarzschild radius [\cite{Peer+07,
    Ghirlanda+13, Iyyani+13, Larsson+15}]. Interestingly, this result
is aligned with recent constraints found by [\cite{VLP13}], that
showed that the conditions for full thermalization takes place only if
dissipation takes place at intermediate radii, $\sim 10^{10}$~cm,
where the outflow Lorentz factor is mild, $\Gamma \sim
10$. Furthermore, this radius of $\sim 10^{8.5}$~cm is a robust radius
where jet collimation shock is observed in numerical simulations
[\cite{ZWM03, MI13}]. These results thus point toward a new
understanding of the early phases of jet dynamics.

A second interesting implication is an indirect way of constraining
the magnetization of the outflow.  It was shown by [\cite{ZM02, DM02,
    ZP09}] that for similar parameters, the photospheric contribution
in highly magnetized outflows is suppressed. Lack of pronounced
thermal component can therefore be used to obtain a lower value on the
magnetization parameter, $\sigma$ [\cite{ZP09}]. Furthermore, it was
recently shown [\cite{BP14}] that in fact in the framework of standard
magnetic reconnection model, conditions for full thermalization do not
exit in the entire region below the photosphere. As a result, the
produced photons are up-scattered, and the resulting peak of the Wien
distribution formed is at $\gsim 10$~MeV. This again leads to the
conclusion that identification of thermal component at energies of
$\lsim 100$~keV must imply that the outflow cannot be highly
magnetized.

\section{Summary and conclusion}
\label{sec:summary}

We are currently in the middle of a very exciting epoch in the study
of GRB prompt emission. Being very short, random and non-repetitive,
study of the prompt emission is notoriously difficult. The fact that
no two GRBs are similar makes it extremely difficult to draw firm
conclusions that are valid for all GRBs. Nonetheless, following the
launch of {\it Swift} and {\it Fermi}, ample observational and
theoretical efforts have been put in understanding the elusive nature
of these complex events. I think that it is fair to say that we
are finally close to understanding the essence of it.

To my opinion, there are two parts to the revolution that take place
in the last few years. The first is the raise of the time-dependent
spectral analysis, which enables a distinction between different
spectral components that show different temporal evolution. A
particularly good example is the temporal behavior of the high energy
(GeV) part of the spectrum, that is lagging behind lower energy
photons.  This temporal distinction enables a separate study of each
component, and points towards more than a single emission zone. This
distinction, in fact, is aligned with the initial assumptions of the
``fireball'' model, in which internal collisions (or several episodes
of magnetic energy dissipation) lead to multiple emission zones.

The second part of the revolution is associated with the
identification of a thermal component on top of the non-thermal
spectra. For many years, until today, the standard fitting of GRB
spectra were, and still are carried using a mathematical function,
namely the ``Band'' model. Being mathematical in nature, this model
does not have any ``preferred'' physical scenario, but its results can
be interpreted in more than one way. As a result, it is difficult to
obtain a theoretical insight using these fits. As was pointed out over
15 years ago, basic radiative models, such as synchrotron, fail to
provide a valid interpretation to the obtained results. Moreover,
while a great advantage of this model is its simplicity, here lies
also its most severe limitation: being very simply, it is not able to
account for many spectral and temporal details, which are likely
crucial in understanding the underlying physics of GRBs.

It was only in recent years, with the abandoning of the ``Band'' model
as a sole model for fitting GRB prompt emission data, that rapid
progress was enabled. The introduction of thermal emission component
played a key role in this revolution. First, it provides a strongly
physically motivated explanation to at least part of the
spectrum. Second, the values of the parameters describing the
non-thermal part of the spectra are different than the values derived
without the addition of a thermal component; this makes it easier to
provide a physical interpretation to the non-thermal part. Third, the
observed well defined temporal behavior opened a new window into
exploring the temporal evolution of the spectra. These observational
realization triggered a wealth of theoretical ideas aimed at
explaining both the observed spectral and temporal behaviors.

Currently, there is still no single theoretical model that is accepted
by the majority of the community. This is due to the fact that
although it is clear that synchrotron emission from optically thin
regions cannot account for the vast majority of GRBs, pure thermal
component is only rarely observed. Furthermore, clearly the very high
energy (GeV band) emission has a non-thermal origin, and therefore
even if thermal component does play an important role, there must be
additional processes contributing to the high energy part. Moreover,
while thermal photons are observed in some GRBs, there are others in
which there is no evidence for such a component. Thus, whatever
theoretical idea may be used to explain the data, it must be able to
explain the diversity observed.

At present epoch, there are three leading suggestions for explaining
the variety of the data. The first is that the variety seen is due to
different in magnetization. It is indeed a very appealing idea, if it
can be proved that the variety of observed spectra depends only on a
single parameter. The second type of models consider the different jet
geometries, and the different observing angles relative to the jet
axis. This is a novel approach, never taken before, and as such there
is ample of room for continuing research in this direction.  The third
type of models consider sub photospheric energy dissipation as a way
of broadening the ``Planck'' spectra. The observed spectra in these
models thus mainly depends on the details of the dissipation process,
and in particular the optical depth in which it takes place.

All of these models hold great promise, as they enable not only to
identify directly the key ingredients that shape the observed spectra,
but also enable one to use observations to directly infer physical
properties. These include the jet dynamics, Lorentz factor, geometry
($\Gamma$ as a function of $r$, $\theta$ and maybe also $\phi$), and
even the magnetization. Knowledge of these quantities thus directly
reflects on answering basic questions of great interest to astronomy,
such as jet launching, composition and collimation.

Thus, to conclude, my view is that we are in the middle of the 'prompt
emission revolution'. It is too early to claim that we fully
understand the prompt emission - indeed, we have reached no consensus
yet about many of the key properties, as is reflected by the large
number of different ideas. However, we understand various key
properties of the prompt emission in a completely different way than
only 5 - 10 years ago. Thus, I believe that another 5- 10 years from
now there is a good chance that we could get to a conclusive idea
about the nature of the prompt emission, and would be able to use it
as a great tool in studying many other important issues, such as
stellar evolution, gravitational waves and cosmic rays.

\acknowledgement
I would like to thank Felix Ryde for numerous number of useful discussions.

\bibliographystyle{aps-nameyear}      



\end{document}